\documentclass[twocolumn,showpacs]{revtex4}

\usepackage{graphicx}
\usepackage{dcolumn}
\usepackage{bm}
\usepackage{hyperref}
\usepackage{subfigure}
\usepackage{cancel}
\usepackage{color}

\begin{document}

\title{Initial data for neutron star binaries with arbitrary spins}

\author{Petr Tsatsin$^1$ and Pedro Marronetti$^{1,2}$}
\affiliation{$^1$ Department of Physics,
	     Florida Atlantic University, 
	     Boca Raton, FL 33431, USA}
\affiliation{$^2$ Division of Physics,
	     National Science Foundation, 
	     Arlington, VA 22230, USA}

\pacs{
04.25.D-,	
04.30.Db,	
95.30.Sf,	
97.60.Jd        
}


%
\newcommand\be{\begin{equation}}
\newcommand\ba{\begin{eqnarray}}

\newcommand\ee{\end{equation}}
\newcommand\ea{\end{eqnarray}}
\newcommand\et{{\it et al.~}}
\newcommand\p{{\partial}}
\newcommand{\rem}[2]{{(\bf {\color{red} #1}}: {\it #2})}

\begin{abstract}

The starting point of any general relativistic numerical simulation is a solution of the Hamiltonian and momentum constraints that (ideally) represents an astrophysically realistic scenario. We present a new method to produce initial data sets for binary neutron stars with arbitrary spins and orbital eccentricities. The method only provides approximate solutions to the constraints. However, we show that the corresponding constraint violations subside after a few orbits, becoming comparable to those found in evolutions of standard conformally flat, helically symmetric binary initial data. We evolve in time data sets corresponding to binaries with spins aligned, zero and anti-aligned with the orbital angular momentum. These simulations show the orbital ``hang-up" effect previously seen in binary black holes. Additionally, they show orbital eccentricities that can be up to one order of magnitude smaller than those found in helically symmetric initial sets evolutions.

\end{abstract}
\maketitle

\section{Introduction}

Binary neutron stars ({\bf BNS}) are currently one of the most studied objects in astrophysics due to their potential as engines for short gamma-ray bursts \cite{Gehrels24082012, Gomboc:2012ff} and as generators of detectable gravitational waves ({\bf GW}) \cite{Sathyaprakash:2009xs}. Recent detection rate estimations for the advanced interferometric detectors are in the range of $0.4-400$ BNS events per year \cite{Abadie:2010cf}, making the observation of GW from BNS very likely in the next few years. However, given the complex nature of NS, numerical modeling of the last few orbits and merger of such binaries is essential for the interpretation of the corresponding GW signatures.

Every numerical simulation has a starting point that is, essentially, a snapshot of all the fields (gravitational, hydrodynamical, electromagnetic, etc.) at a given time. Depending on the characteristics of the modeling formalism, these fields can either be freely specified or constrained by set of conditions. Numerical simulations in general relativity that are based on ``3+1"-type formalisms are of the latter kind: the fields have to be solutions of the Hamiltonian and momentum constraints to be consistent with the full set of the Einstein field equations \footnote{Other constraints could also be present if additional fields (such as electromagnetic) are included.}. The Hamiltonian and momentum constraints are four coupled second order elliptic PDEs that are solved numerically through some iterative procedure that starts with an initial guess and loops around the equations, correcting the fields until some predetermined convergence criteria is reached. Since these four equations are not enough to determine the ten independent components of the spacetime metric, the modeler has the freedom to choose additional constraints/conditions. When it comes to finding initial states for BNS in circular orbits, the most popular approach is the Wilson-Mathews conformal ``thin-sandwich" scheme \cite{Wilson95, Wilson:1996ty} which consists of restricting the solutions with three extra conditions: that the spatial $3$-metric $\gamma_{ij}$ be conformally flat, that the slicing be maximal ($tr(K_{ij})=0$, where $K_{ij}$ is extrinsic curvature), and that the spacetime be helically symmetric (simply put, that the fields be time independent in the frame that corotates with the binary). The first two conditions reduce the number of unknowns from ten to five (the conformal factor, the lapse function and the three components of the shift vector). The third, however, is related to another concern of the initial data ({\bf ID}): the need for it to represent ``astrophysically realistic" scenarios. 

A great deal could be discussed about what constitutes an astrophysically realistic BNS; particularly since we are largely ignorant of the state of matter inside a neutron star. However, there are two aspects of these systems that most researchers agree on. One is related to the circularity of the orbits of a BNS system that has evolved in isolation. In cases like these, it is expected that any initial eccentricity the binary may have acquired at birth will be minimized by emission of gravitational radiation \cite{Peters:1963ux} (for instance, the BNS known as PSR $1913+16$ is supposed to end its life with eccentricities of about $10^{-6}$ \cite{Oslowski:2009zr}). Recent estimates of Advanced LIGO event rates indicate that the fraction of BNS formed by stellar evolution of binary systems with eccentricities larger than $0.01$ is, for the most optimistic scenario, below $2\%$ \cite{Kowalska:2010qg}. BNS formed by dynamical capture are expected to have higher eccentricities by the time they merge. However, their detection event rate is uncertain and likely to be much smaller than the one corresponding to the binary evolution channel (see \cite{Gold:2011df} and references therein). While the helical symmetry condition demands exact circular orbits, it actually produces non-negligible eccentricities that, as it is shown below, surpasses $0.01$ by a factor of several (see also \cite{Miller03c}).

The other aspect of astrophysically realistic ID sets is that they should, in principle, be able to describe spinning stars. Finding ways to construct ID for binaries with spinning NS has been a problem more difficult to address. Neutron stars obey hydrodynamical equations such as the general relativistic versions of the continuity and Euler equations and any description of fluids in rotation should be consistent with them. The original work of Wilson and collaborators handled the hydrodynamics through a lengthy evolution process that, while not practical for the production of ID sets, permitted the imposition of arbitrary NS spins by the way of angular momentum drivers that forced the stars to adopt the desired rotations. The impractical nature of this method gave rise to a search for simpler techniques. One of the first to be considered was the special case when the two stars are tidally locked or ``corotating" \cite{Baumgarte:1997xi, Baumgarte98c}. Under corotation, the fluid is static in the frame that rotates with the BNS and the continuity equation is trivially satisfied. These cases are unlikely to exist in nature since they would require fluid viscosities unrealistically high \cite{Bildsten92, Kochanek92} but they are still useful to test new algorithms and numerical codes. Corotating solutions were followed by solutions with null fluid vorticity (``irrotational"). This formalism was developed to find solutions for BNS with (nearly) zero spin by the way of specifying the fluid velocity as the gradient of a potential \cite{Bonazzola97, PhysRevD.57.7292, Teukolsky98, Shibata98}. This potential is obtained from an additional elliptic equation derived from imposing zero vorticity. Since its introduction, this formalism has become the preferred method for BNS ID production and several groups have developed codes and techniques for its implementation \cite{Bonazzola97, Marronetti:1998xv, Bonazzola:1998yq, Marronetti:1999ya, Uryu:1999uu, Uryu00a, Gourgoulhon:2000nn, Taniguchi:2001qv, Taniguchi:2002ns, Kiuchi:2009jt}. All simulations of BNS in circular orbits performed to date are either based on corotating or irrotational ID sets \cite{Faber:2012rw}. In recent times, several groups have experimented with non-conformally flat techniques \cite{Uryu:2005vv, PhysRevD.78.104016, Uryu:2009ye}. Among these, the works by Anderson \et \cite{Anderson:2007kz}, Gold \et \cite{Gold:2011df}, East \et \cite{East:2012zn} and Kastaun \et \cite{Kastaun:2013mv} are of relevance to this paper and more about them is said in section \ref{ID_sets}.

However, in general, neutron stars in binaries are expected to be spinning. A good example of this is the double pulsar PSR J0737-3039 \cite{Lyne:2004cj} that could have one of the stars spinning at a rate of about $\sim 27$ms at the time of the merger \cite{Tichy:2011gw}. Numerical schemes to produce ID for spinning BNS have been presented by Marronetti and Shapiro \cite{Marronetti:2003gk}, Baumgarte and Shapiro \cite{Baumgarte:2009fw} and Tichy \cite{Tichy:2011gw, Tichy:2012rp}. All of these are based on the Wilson-Mathews helically symmetric, conformally flat approximation. Like in the case of irrotational BNS, these methods also rely on advanced computationally intensive iterative algorithms.

We present here a method for producing ID corresponding to spinning BNS that also allows for arbitrary orbital and radial velocities. This freedom gives more control over the orbital eccentricity than in the case of helically symmetric methods. Our method does not look for solutions of the Hamiltonian and momentum constraints: their satisfaction is only asymptotic with binary separation. This has the advantage of not requiring the numerical solution of elliptic equations, thus greatly simplifying its implementation. To find out the impact of the resulting constraint violations, we produced non-spinning BNS ID sets, evolved them in time, and compare the results with those of simulations starting with irrotational ID generated with the LORENE library \cite{Gourgoulhon:2000nn, LORENE_web}. To facilitate the simulations, we implemented our method as a module (``thorn") of the Einstein Toolkit (ET) \cite{ET_web, ET2011}. We evolved these BNS for up to seven orbits before merger and show that, for the grid resolutions and binary separations studied here, the constraint violations in both simulations become comparable before the merger \footnote{During the preparation of this article, Kastaun \et \cite{Kastaun:2013mv} and Alic \et \cite{Alic:2013xsa} confirmed that the constraint violations' reduction becomes more dramatic when evolving the sets with the CCZ4 formulation instead of BSSNOK.}. Additionally, we show that our ID sets can lead to orbits that exhibit eccentricities smaller than those resulting from evolving helically symmetric ID sets. Finally, we present the evolution of ID sets with spinning NS (spins aligned and counter-aligned to the orbital angular momentum) that showcase the ability of our method to handle rotating stars. These simulations present the orbit ``hang-up" effect \cite{Campanelli:2006uy} in BNS that is also seen in \cite{Kastaun:2013mv}.

This article is organized as follows. Sections \ref{ID_sets} and \ref{tests} introduce our method and present some tests respectively. In section \ref{results} we use our approach to construct ID for non-spinning and spinning binaries and show the results of their evolution in full general relativistic hydrodynamics using the Einstein Toolkit \cite{ET_web}. Finally, we will briefly summarize our findings in Sec.\ref{conclusions}. Several of the figures presented below contain temporal and spatial coordinates which are displayed both in SI units and normalized by a nominal constant $M_3 \equiv 3 M_{\odot}$.

\section{Construction of Initial Data sets}
\label{ID_sets}

\subsection*{Single rotating neutron stars}
  
Our approximation to ID for BNS starts with the solution of an isolated rotating NS in equilibrium. The study of stationary rotating NS has been undertaken by a number of groups in the past (see \cite{lrr-2003-3} and therein). These studies have led to the creation of publicly available codes specially designed to find numerical solutions in the framework of the theory of general relativity. One of such codes is {\bf RNS}, developed by Stergioulas and Friedman \cite{Stergioulas95, RNS_web} and based on the Komatsu-Eriguchi-Hachisu method \cite{Komatsu89, Cook1994ApJ} which describes the geometry of stationary and axisymmetric rotating NS with a metric of the form
  \ba
    ds^2 &=& -e^{A+B}dt^2 + e^{2 C}(dr^2+r^2d\theta^2) \nonumber \\
         &&+e^{A-B}r^2\sin^2\theta ~(d\phi-D ~dt)^2,
    \label{eq1}
  \ea
where the metric functions $A$, $B$, $C$ and $D$ depend on $r$ and $\theta$. The equations for the gravitational and matter fields are then solved using a combination of integral and finite-differencing techniques \cite{Cook1994ApJ}.

\subsection*{Our initial data}
Our ID is produced following these three steps:\\

\indent (i)  Calculate the fields corresponding to two isolated rotating neutron stars using RNS,\\
\indent (ii) Rotate (if needed) and boost independently each solution and map them into a single inertial frame, making sure that the BNS total linear momentum is zero, and\\
\indent (iii) Superpose the fields as indicated below.\\

Step (i) is straightforward and generates two stationary solutions for rotating NS. Each one is originally given in the reference frame of the RNS code: $x'^\mu$. RNS provides solutions in polar coordinates with the NS rotating around the $z'$ axis. Since we are interested in NS with spins arbitrarily aligned, a rotation of the solutions may be required in combination with the boost. To simplify the notation, we will ignore here the rotation (i.e., we will consider BNS with the spins in the direction of the orbital angular momentum). 

Step (ii) is inspired by standard binary black hole ({\bf BBH}) superposition methods \cite{Matzner98a, Marronetti00a} and starts with the Lorentz boost of the RNS coordinates into the Cartesian inertial frame coordinates $x^\mu$
\be
      x^{\nu}= {_a\Lambda^{\nu}_{\mu}} x'^{\mu},
\label{original_LT}
\ee
where the Lorentz transformation $_a\Lambda^{\nu}_{\mu}$ is a function of the corresponding boost velocity ${\bf v}_a$, with the index $a$ labeling each star ($a=1,2$). We now map the metric functions to the new coordinate system using (\ref{original_LT})
\ba
       _aA'(x'^{\mu}) &\rightarrow & _aA'(x^{\nu})  \nonumber \\
       _aB'(x'^{\mu}) &\rightarrow & _aB'(x^{\nu})  \nonumber \\
       _aC'(x'^{\mu}) &\rightarrow & _aC'(x^{\nu})  \nonumber \\
       _aD'(x'^{\mu}) &\rightarrow & _aD'(x^{\nu}). \nonumber 
\ea
This allows us to write the spacetime metric $g'^a_{\tau\lambda}$ as a function of the new coordinates $x^{\nu}$ and Lorentz transform it to the inertial frame
\be
       {_ag_{\tau\lambda}(x^{\nu})} = {}_a\Lambda^{\delta}_{\tau} {}_a\Lambda^{\eta}_{\lambda} ~{_ag'_{\delta\eta}(x^{\nu})}, \label{metric_LT}
\ee
where there is no summation over {\it a}. Since we are interested in solutions in terms of a ``3+1" decomposition of the metric, we extract from ${_ag_{\tau\lambda}(x^{\nu})}$ the corresponding lapse function $_a\alpha$, shift vector $_a\beta^i$, and spatial metric $_a\gamma_{ij}$. Similarly, a mapping/transformation is applied to the rest mass and fluid velocity $_a\rho'$ and $_av'^i$, to obtain the fields $_a\rho$ and $_av^i$. The latter is defined as $v^i = (u^i/u^0 + \beta^i)/\alpha$, where $u^\mu$ is the 4-velocity of the fluid.

Step (iii) constructs a single global set of gravitational fields by a superposition of these two solutions
\ba
     \alpha  &=& _1\alpha + {}_2\alpha-1  \nonumber \\                          
     \beta^i &=& _1\beta^i + {}_2\beta^i  \nonumber \\
     \gamma_{ij} &=& _1\gamma_{ij} + {}_2\gamma_{ij}-\delta_{ij}.   \label{ADMfields}
\ea
Since there is no overlap between the stars, the superposition of the hydrodynamics fields is simply 
\ba
     \rho &=& _1\rho + {}_2\rho \nonumber \\
     v^i  &=& _1v^i + {}_2v^i.
     \label{originalHydrofields}       
\ea
In addition to the stellar matter, a pervasive atmosphere is added outside the stars with a relatively low density $\rho_{atm}=10^{-7}\rho_{max}(t=0)$ and zero velocity. The remaining hydrodynamical fields (pressure and internal energy) are set by the equation of state (EOS). Finally, the extrinsic curvature is calculated using 
\be
     K_{ij}=-\frac{1}{2\alpha}\left(\partial_t \gamma_{ij} - \mathcal{L}_{\beta}\gamma_{ij}\right),
     \label{Kij}
\ee      
where $\mathcal{L}_{\beta}$ is the Lie derivative along the direction of the shift vector.

Superpositions such as this have been used for BNS simulations in the past. We will refer to them as ``simple superposition". In particular, Anderson \et \cite{Anderson:2007kz}, Gold \et \cite{Gold:2011df} and, more recently, East \et \cite{East:2012zn} and Kastaun \et \cite{Kastaun:2013mv} have generated and evolved BNS ID sets using either a superposition similar to the one described above or even adding the extra step of actually solving the Hamiltonian and momentum constraints \footnote{This is achieved by the use of a conformal factor as an auxiliary field derived from solving the Hamiltonian constraint, and by corrections to the superposed shift vector that solve the momentum constraint \cite{East:2012zn}.}. However, ID sets constructed in this way present two undesirable features which are evident during their time evolution: relatively large oscillations of the stellar shape and orbital eccentricities. Below we describe two modifications to simple superposition that reduce spurious effects. Both of them play a role in controlling shape oscillations and eccentricities, even when each modification plays a dominant role in controlling a particular problem.

\begin{figure}[h]
    \includegraphics[width=0.5\textwidth]{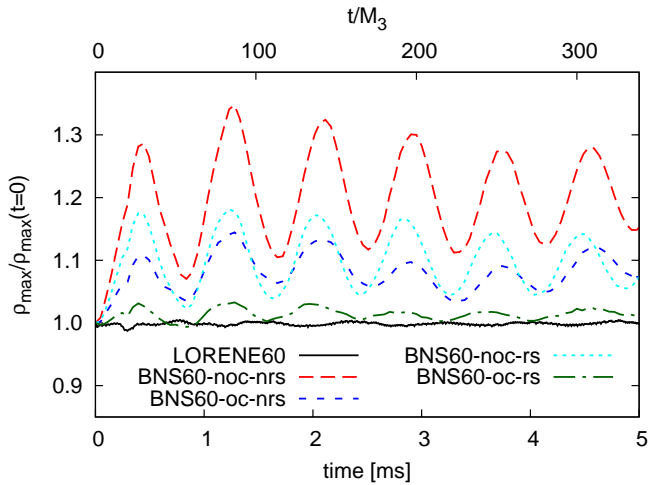}
    \caption{Maximum rest mass density for simulations of non-spinning binaries starting at a coordinate separation of $60$ km. The evolution of a reference ID set ({\it LORENE60} in Table \ref{Table1}) is compared with four ID sets constructed using superposition ({\it BNS60-xxx-yyy}). The latter cover cases with (oc) and without (noc) ``over-contraction" (Eq. (\ref{n_LT}) with $n$ equal to 4 or 1 respectively) and with (rs) and without (nrs) rescaling (Eq. (\ref{Hydrofields})). The value $M_3$ used for normalization in the upper $x$-axis is defined as $3 M_{\odot}$.} 
\label{over_contraction}
\end{figure}

Oscillations of the stellar shape that can be observed by monitoring the central rest mass density, as shown in Fig. \ref{over_contraction}. The solid line corresponds to the evolution of a reference ID with initial separation of $60$ km generated by the LORENE library ({\it LORENE60} in Table \ref{Table1}, also known as $\mathtt{G2\_I12vs12\_D5R33\_60km}$). Details of the time evolution are given in section \ref{tests}. The curve labeled {\it BNS60-noc-nrs} corresponds to a comparable ID set generated using the steps mentioned above. This curve presents large oscillations that contrast with the flatness of the evolution of {\it LORENE60}. While the amplitude of the oscillations is partially due to the grid sparseness, the difference between both runs indicates that there are additional causes at play. One of them is the fact that simple superposition does not make any attempt at coupling the hydrodynamical fields (obtained from single NS solutions) with the metric fields calculated in (\ref{ADMfields}). These oscillations will diminish with increasing binary separation (see Table I in \cite{East:2012zn}) and could, to some extent, be controlled by introducing fluid viscosity terms in the Euler equations. However, we explored alternative ways of minimizing this effect that could be easily implemented in the ID generation code. Visual inspection of the stellar cross-section on the orbital plane showed that the stars corresponding to the LORENE ID set are approximately oval in shape, with the diameter along the direction between stellar centers larger than the one in the direction of the orbital velocity. While our ID also presents this feature (a result of the Lorenz contraction in the boost direction), it does so at a lesser degree. As an experiment, we tried increasing the deformation of the stars in our ID by replacing the coordinate transformation (\ref{original_LT}) with
\be
      x^{\nu}= ({_a\Lambda^{\nu}_{\mu}})^n x'^{\mu},
\label{n_LT}
\ee
where the exponent indicates that transformation matrix was applied $n$ times. The result of this experiment corresponding to $n=4$ is shown as the evolution of the ID set {\it BNS60-oc-nrs} in Fig. \ref{over_contraction}.
One problem of this method is that the free parameter {\it n} is confined to integer values, limiting the method's fine-tuning capabilities. To compensate for this, we devised an alternative that achieves over-correction by using a single Lorentz transformation
\be
      x^{\nu}= ~{_{fa}\Lambda^{\nu}_{\mu}} x'^{\mu}, 
\label{f_LT}
\ee
where ${_{fa}\Lambda^{\nu}_{\mu}}$ is now a function of a velocity ${\bf v}_{fa} = f \times {\bf v}_a$ with $f$ a real positive number. A comparison of the results obtained with these two techniques is given in Fig. \ref{Comp_n_ocf} which is discussed in section \ref{varying_res}. Note that the over-contraction is applied only in the coordinate transformation (\ref{n_LT}) or the one in (\ref{f_LT}) and {\it not} in the transformation of the components of the metric (\ref{metric_LT}) or the hydrodynamical fields.

The second problem of the ID recipe as given at the beginning of the section is related to the orbital eccentricity: using the superposition outlined in Eqs. (\ref{ADMfields}, \ref{originalHydrofields}) leads to binaries that exhibit non-circular orbits. As expected, this problem also diminishes with increasing binary separation. 
\begin{figure}[h]
    \includegraphics[width=0.5\textwidth]{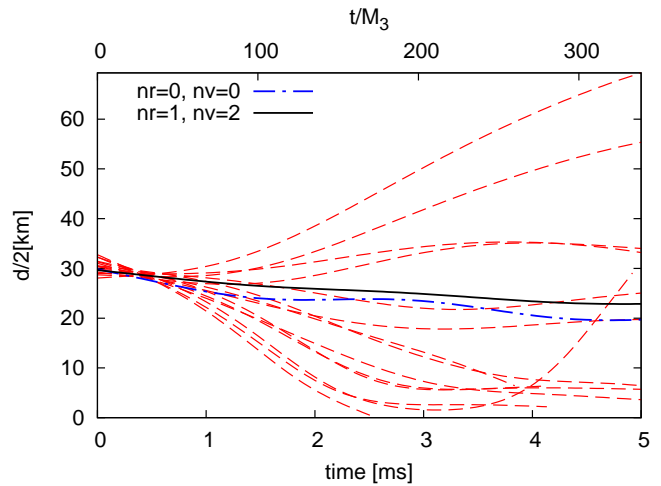}
    \caption{Coordinate separation for simulations of non-spinning binaries starting at a distance of $60$ km. The effect of the rescaling of hydrodynamical fields is shown for all the integer values of the exponents $nr$ and $nv$ in Eq. (\ref{Hydrofields}) from $0$ to $3$. The lowest eccentricity is achieved by the combination ($nr=1$, $nv=2$) depicted with a solid curve. All these runs employed the over-contraction of Eq. (\ref{n_LT}) with $n=4$.}
\label{rescaling_tests}
\end{figure}
We believe this is primarily due to the fact that hydrodynamical fields are not ``adjusted" to reflect the change in the gravitational fields caused by superposition. For instance, the gravitational field resulting from superposition is stronger than that of a single star, leading to a more compact stellar structure. Again we experimented with alternatives to the simple recipe given in Eqs. (\ref{originalHydrofields}). One way to compensate for the increase in stellar density is to modify the RNS profiles $_a\rho$ and $_av^i$ with factors dependent on some of the gravitational fields. A simple choice, albeit not the only one, is to use $(\alpha/_a\alpha)^{nr}$ for $\rho$ and $(_a\alpha/\alpha)^{nv}$ for $v^i$, where the integer exponents $nr$ and $nv$ are free parameters:
\ba
     \rho &=& \left(\frac{\alpha}{_1\alpha}\right)^{nr} {_1}\rho  
            + \left(\frac{\alpha}{_2\alpha}\right)^{nr} {_2}\rho  \nonumber \\
     v^i  &=& \left(\frac{_1\alpha}{\alpha}\right)^{nv} {_1}v^i 
            + \left(\frac{_2\alpha}{\alpha}\right)^{nv} {_2}v^i.
     \label{Hydrofields}       
\ea
This rescaling could be interpreted as a weighted average of the hydrodynamics fields by a measure of the spacetime curvature. We tested this formula for different values of the exponents ranging from zero to $3$ and the results are presented in Fig. \ref{rescaling_tests} \footnote{The use of negative exponents is quickly ruled out since they modify the matter fields in the ``wrong" direction.}. The simulation that exhibits the smaller orbital eccentricity corresponds to the case $nr=1$ and $nv=2$ and, based on these tests, we decided to adopt those values of all the runs in this article that employ rescaling. 

\begin{figure}[h]
    \includegraphics[width=0.5\textwidth]{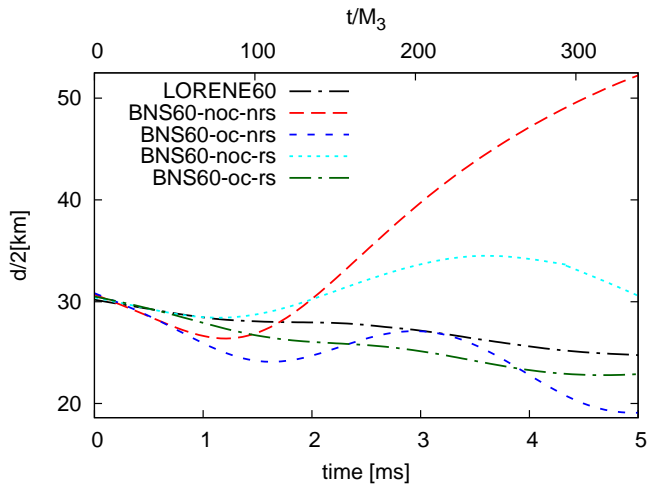}
    \caption{Coordinate separation for the runs of Fig. \ref{over_contraction}.}
\label{superposition_rescaling}
\end{figure}

Each one of the modifications to superposition introduced here (over-contraction and rescaling) affect both shape oscillations and orbital eccentricity. Figures \ref{over_contraction} and \ref{superposition_rescaling} show this by comparing runs with sets with (oc) and without (noc) over-contraction and with (rs) and without (nrs) rescaling. It is clear that optimal results are obtained when both modifications are applied (curves labeled {\it BNS60-oc-rs}).

To summarize, our recipe for BNS ID consists in following steps (i - iii), employing either the coordinate transformation given in Eq. (\ref{n_LT}) or the one in (\ref{f_LT}) and the calculation of the fields given in Eqs. (\ref{ADMfields}), (\ref{Kij}), and (\ref{Hydrofields}).

We conclude this section with some thoughts about the motivation behind the over-contraction and rescaling techniques. We do not offer here any rigorous justification for them beyond their empirical success. While we believe that proper mathematical studies could provide strong reasons for the choices we made and suggest improvements, the process by which we constructed these methods was a combination of intuition and trial-and-error. As we already mentioned, we believe that the uncoupling of the matter and gravitational fields caused by simple superposition is the root of the problem and over-contraction and rescaling constitute a rather simple attempt at adjusting the matter fields to the new gravitational background. It has been suggested to us that other alternatives could be used to attain similar goals. One, for instance, is the use of different coordinate systems including variations of the lapse and shift that, if properly chosen, would have the additional advantage of not exacerbating the constraints' violations. We have not tried this here since coordinate transformations would not affect the relation between gravitational and hydrodynamical fields and, in our opinion, they do not address directly the problem. However, we do believe that our methods can be improved and we will dedicate future studies to that effect.

\section{Numerical Tests}
\label{tests}
\begin{table*}
  \begin{tabular}{lcccccccccccc}
    \hline
              & $d/M^*$& $M_0[M_{\odot}]$ & $M_{ADM}[M_{\odot}]$ & $J_{ADM}[M_{\odot}^2]$ & $a$ & $v^\phi$/c & $v^r/c$ & n/f & rl & set & res\\
       \hline
   BNS60\_noc\_nrs & 13.45 & 3.566 & 3.311 & 9.377 & 0 & 0.1450 & 0 & 1 & 5 & (0.0,1.0,0.01) & $0.50M_\odot$\\
   BNS60\_oc\_nrs  & 13.45 & 3.454 & 3.207 & 9.081 & 0 & 0.1450 & 0 & 4 & 5 & (0.0,1.0,0.01) & $0.50M_\odot$\\
   BNS60\_noc\_rs  & 13.45 & 3.385 & 3.138 & 9.849 & 0 & 0.1450 & 0 & 1 & 5 & (0.0,1.0,0.01) & $0.50M_\odot$\\
   BNS60\_oc\_rs   & 13.45 & 3.279 & 3.039 & 9.537 & 0 & 0.1450 & 0 & 4 & 5 & (0.0,1.0,0.01) & $0.50M_\odot$\\
       \hline 
   LORENE60   & 13.45 & 3.250 & 3.005 &  9.716 & 0 & 0.1238 &  0.0    &   - & 5 & (0.0,1.0,0.01) & $0.50M_\odot$\\ 
   LORENE80   & 18.02 & 3.250 & 3.011 & 10.825 & 0 & 0.1098 &  0.0    &   - & 5 & (0.0,1.0,0.01) & $0.50M_\odot$\\
   BNS50n     & 11.30 & 3.250 & 3.006 &  8.985 & 0 & 0.1615 & -0.0010 &   4 & 5 & (0.0,1.0,0.01) & $0.50M_\odot$\\
   BNS60n     & 13.45 & 3.250 & 3.020 &  9.718 & 0 & 0.1450 & -0.0040 &   4 & 5 & (0.0,1.0,0.01) & $0.50M_\odot$\\
   BNS70n     & 15.82 & 3.250 & 3.018 & 10.060 & 0 & 0.1360 & -0.0045 &   4 & 5 & (0.0,1.0,0.01) & $0.50M_\odot$\\
   BNS80n     & 18.02 & 3.250 & 3.020 & 10.497 & 0 & 0.1265 & -0.0050 &   4 & 5 & (0.0,1.0,0.01) & $0.50M_\odot$\\
       \hline
   LORENE60\_lr & 13.45 & 3.250 & 3.004 & 9.716 & 0 & 0.1238 &  0.0    &   - & 5 & (1.0,0.5,0.10) & $0.50M_\odot$\\
   LORENE60\_mr & 13.45 & 3.250 & 3.005 & 9.716 & 0 & 0.1238 &  0.0    &   - & 6 & (1.0,0.5,0.10) & $0.25M_\odot$\\
   LORENE60\_hr & 13.45 & 3.250 & 3.005 & 9.716 & 0 & 0.1238 &  0.0    &   - & 6 & (1.0,0.5,0.10) & $0.1875M_\odot$\\
   BNS60n\_lr   & 13.45 & 3.252 & 3.018 & 9.718 & 0 & 0.1490 & -0.0020 & 2.1 & 5 & (1.0,0.5,0.10) & $0.50M_\odot$\\
   BNS60n\_mr   & 13.45 & 3.252 & 3.018 & 9.718 & 0 & 0.1490 & -0.0010 & 2.1 & 6 & (1.0,0.5,0.10) & $0.25M_\odot$\\
   BNS60n\_hr   & 13.45 & 3.252 & 3.018 & 9.718 & 0 & 0.1490 & -0.0010 & 2.1 & 6 & (1.0,0.5,0.10) & $0.1875M_\odot$\\
       \hline   
   BNS80u     & 18.02 & 3.250 & 3.034 & 12.142 &  0.327 & 0.1250 & -0.0050 & 4 & 5 & (0.0,1.0,0.01) & $0.50M_\odot$\\
   BNS80d     & 18.02 & 3.250 & 3.032 &  8.984 & -0.327 & 0.1265 & -0.0050 & 4 & 5 & (0.0,1.0,0.01) & $0.50M_\odot$\\
   \hline  
  \end{tabular}
  \caption{Parameters of the ID sets used in the binary simulations. The coordinate separation is given by $d$, while $M_0$, $M_{ADM}$ and $J_{ADM}$ are the total rest mass, ADM mass and ADM angular momentum respectively. $M^*$ is the ADM mass of the corresponding LORENE set. $a$ is the dimensionless spin parameter for each NS, and $v^\phi$ and $v^r$ are the tangential and radial components of the boost velocity. $n/f$ is the over-contraction parameter: integers correspond to values of $n$ in (\ref{n_LT}), while real numbers correspond to values of the $f$ in (\ref{f_LT}). $set$ refers to the parameters $\alpha_d$, $\beta_d$ and $\epsilon_{diss}$ used in the evolution and $res$ is the maximum spatial resolution. The numerals in the labels indicate the initial coordinate separation in km. The suffixes of the top part of the table indicate whether over-correction was applied (oc) or not (noc) and whether rescaling was applied (rs) or not (nrs). The suffixes of the bottom part of the table indicate if the stellar spins are up (u), down (d) or null (n) and if the simulation was performed in low (lr), medium (mr) or high resolution (hr). \label{Table1}}
\end{table*}

The algorithms described in the previous section have been implemented in a numerical module (``thorn") for the ET framework \cite{ET_web, ET2011}. This thorn will be made publicly available in the near future. The code accepts the selection of one or two NS with arbitrary boost velocities, spin directions and initial positions. The characteristics of the NS (mass, spin, EOS) are selected through free parameters in RNS which provides the stellar profiles to be mapped into the ET grid.
To test our code, we generated ID for single and binary NS and evolved them using ET. Almost all the simulations have the same grid domain: 5 levels of mesh refinement, provided by \texttt{Carpet} thorn \cite{Schnetter04}, with box sizes $320M_{\odot}$ (outer boundary), $120M_{\odot}$, $60M_{\odot}$, $30M_{\odot}$ and $15M_{\odot}$, resulting in a resolution of approximately 36 points across the stellar diameter. Since it is not our intention to produce high quality BNS models but to test the viability of our ID, we opted for a computationally affordable and expedient numerical setup. The only exceptions are the medium and high resolution cases presented in the third block of Table \ref{Table1} that have extra levels of mesh refinement, increasing the number of grid points across the star to 72 and 96 respectively.

The BNS simulations for this article make use of xy-plane reflection and $\pi$-rotation symmetries, since they pertain to systems with identical NS with non-precessing orbits. We employ a polytropic EOS ($p=K\rho^{\Gamma}$) with $K=123.6$ and $\Gamma=2.0$ for both the ID sets and the evolution. Spacetime evolution is obtained through the BSSNOK formalism \cite{Nakamura87, Shibata95, Baumgarte:1998te}, provided by the \texttt{McLachlan/ML\_BSSN} thorn \cite{Brown:2008sb}. The general relativistic hydrodynamic equations are evolved by the \texttt{GRHydro} thorn \cite{Baiotti04a}. We use a Marquina Riemann solver with PPM reconstruction of the primitive variables. Kreiss-Oliger dissipation is added to the right-hand-sides of the BSSNOK evolution equations, modulated by the dissipation strength parameter $\epsilon_{diss}$. Finally, the lapse function and shift vector are evolved in time using:
\ba
\partial_t\alpha - \beta^j\partial_j\alpha &=& -2 \alpha ~(\mathrm{tr}(K_{ij})+\mathrm{\alpha_d} ~(\alpha-1)),\nonumber\\
\partial_t\beta^i - \beta^j\partial_j\beta^i &=& 0.75\left(\tilde{\Gamma}^i -\mathrm{\beta_d} \beta^i\right).
\ea
The parameters $\alpha_d$ (``Alpha Driver"), $\beta_d$ (``Beta Driver") and $\epsilon_{diss}$ adopted the values detailed in Table \ref{Table1}.

The runs presented here were performed on the Cray XT5 system ``Kraken" of The National Institute for Computational Sciences (NICS). The computational resources needed for a given simulation are a function of the separation distance and the grid structure, as well as the physical characteristics of the BNS (masses, spins, etc.). The low resolution ($\Delta x=0.50M_{\odot}$) runs for non-spinning BNS starting at a separation of $60$ ($80$) km required close to $2$ ($3.5$) KSUs (thousands of Service Units). However, we need to add to this amount the computer time spent determining the optimal boost velocity. This typically involved ten iterations of about three orbits each, increasing the number of KSUs from $2$ ($3.5$) to $11$ ($20$) for $60$ ($80$) km separation runs. BNS with non-zero spins such as those described in the last block of Table \ref{Table1} will have inspirals of different lengths, requiring computer times that in our cases ranged from $19$ ({\it BNS80d}) to $23$ ({\it BNS80u}) KSUs. The non-spinning BNS simulations with medium and high resolutions described in the third block of Table \ref{Table1} take as much as $105$ and $570$ KSUs respectively (these figures include the determination of the boost velocity). The total amount of computer time spent on the runs presented on this article was close to 1 MSU. The low resolution runs were executed using either 12 or 24 cores. For the medium and high resolution runs, we found that the optimal number of cores was 64 and 240 correspondingly.

\subsection*{Isolated Stars}

We start our tests with four single NS cases: with and without spin, and stationary and boosted. While these seem rather trivial since they only involve mapped solutions of the RNS code into the ET grid, they fulfill two purposes: to test the mapping/boost algorithms and to provide an apt reference in terms of the constraint violations that will be useful when analyzing the simulations of section \ref{results}.
\begin{figure}[h]
     \includegraphics[width=0.5\textwidth]{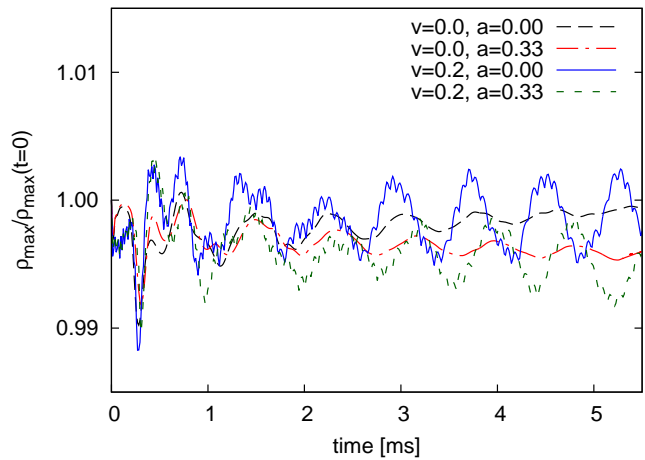}
     \caption{Normalized maximum rest mass density for four single NS test  
        cases that include stationary, spinning and boosted stars.
     \label{rho_sst}}
\end{figure}
\begin{figure}[h]
     \includegraphics[width=0.5\textwidth, keepaspectratio]{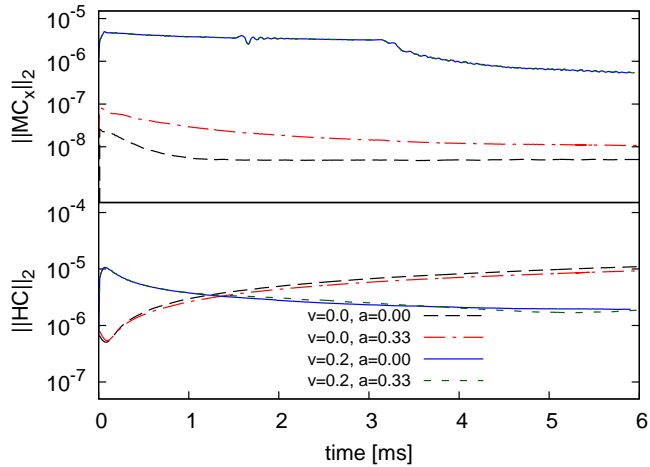}
     \caption{Hamiltonian and momentum constraint violation ($L_2$ norm) for 
        the runs of Fig. \ref{rho_sst}.
     \label{constraints_sst}}
\end{figure}
The NS rest masses are $M_{0}=1.625M_{\odot}$ (compaction ratio = 0.14). The dimensionless spin parameter $a$ ($J/M^2$) is either zero or $0.33$ and the boost velocity $v$ is either zero or $0.2c$. Figure \ref{rho_sst} shows the evolution of the maximum rest mass density. We observe a stable evolution for all four cases, with central density oscillations consistent with our grid resolution. Figure \ref{constraints_sst} shows the Hamiltonian and momentum constraint violations for the four cases under consideration.

\subsection*{Binaries}

\begin{figure}[h]
        \includegraphics[width=0.5\textwidth,keepaspectratio]{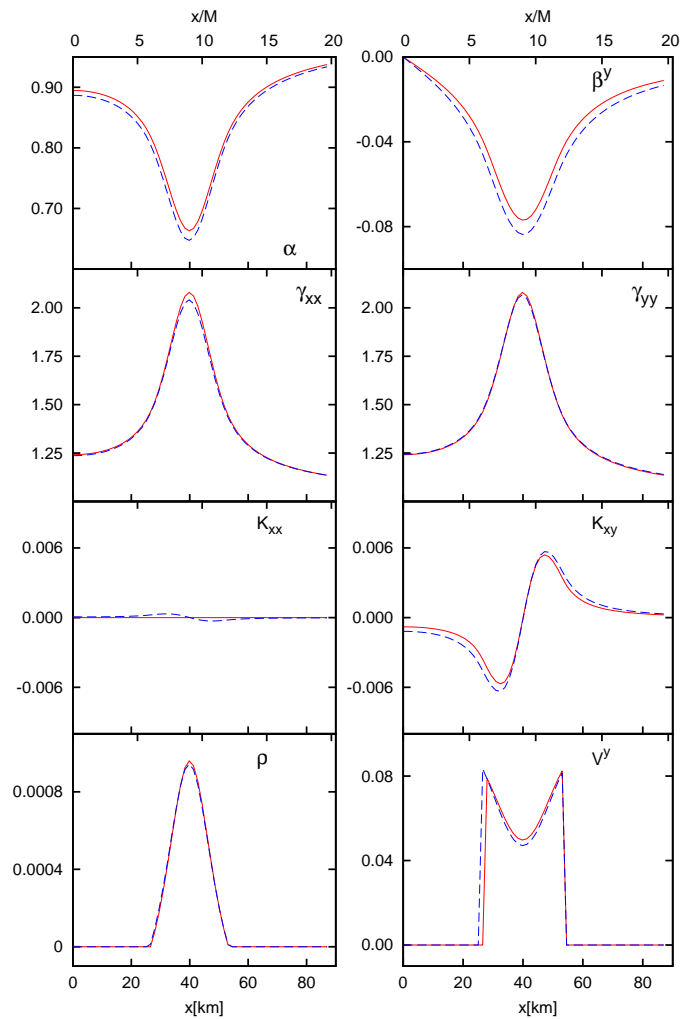}
        \caption{Comparison of several fields from the reference ID set {\it LORENE80} (red solid line) and our non-spinning ID set {\it BNS80n} (blue dashed line). The fields are plotted along the line connecting the two NS centers. 
\label{ID_comp}}
\end{figure}

Figure \ref{ID_comp} presents a comparison of several metric fields between a LORENE-generated irrotational ID ({\it LORENE80} in Table \ref{Table1}, also known as $\mathtt{G2\_I12vs12\_D5R33\_80km}$) and a non-spinning binary ID set with coordinate separation of $80$ km produced with our method ({\it BNS80n} in Table \ref{Table1}). These plots show that even the simple superposition of Eq. (\ref{ADMfields}) gives good agreement. We experimented with higher order alternatives that, while reducing the separation between curves, did not improve noticeably the results of the time evolution. The agreement in the fluid velocity can be further improved by adding small positive spins to the NS. We decided to neglect this higher order correction in this paper.
\begin{figure}[ht]
\centering
\includegraphics[width=0.5\textwidth,keepaspectratio]{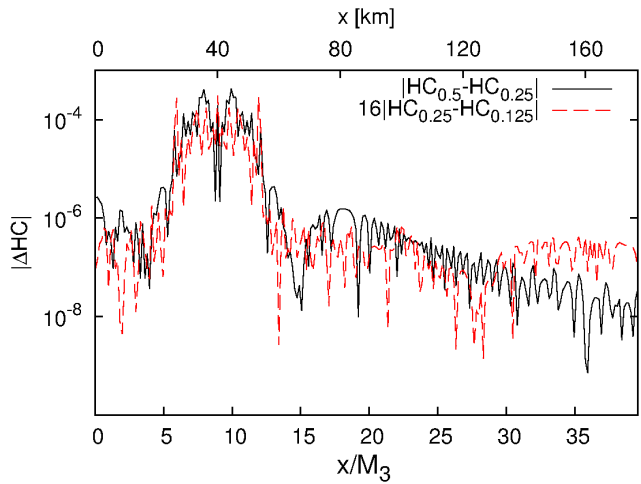}
\caption{Hamiltonian constraint convergence of non-spinning {\it BNS} ID sets with a separation of $80$ km. The plot shows the difference between medium $(0.25M_\odot)$ and low $(0.5M_\odot)$ resolution (solid line), and the difference (scaled for 4th order convergence) between the high $(0.125M_\odot)$ and medium resolution (dashed line).}
\label{HC_CT}
\end{figure}
Figure \ref{HC_CT} shows a point-by-point convergence of the {\it BNS80n} ID set along the line 
connecting the stars that is consistent with the fourth-order of the finite-difference stencils used for spatial derivatives. For this convergence plot, we set the initial data on an extended unigrid for the resolutions $\Delta x=(0.50M_{\odot}, 0.25M_{\odot}, 0.125M_{\odot})$ which correspond to $~(0.738, 0.369, 0.184)$ in km. Almost all our simulations have a central resolution equal to the lowest of the ones shown in Fig. \ref{HC_CT} ($\Delta x=0.50M_{\odot}$). The exception are the medium and high resolution runs presented in the third block of Table \ref{Table1}.

\section{Binary Evolutions}
\label{results}

\subsection{Comparison with known initial data}
\label{comparing_ID}

One of the most important tests of any ID set is provided by its evolution. Qualitatively good agreement between the fields such as the one presented in Fig. \ref{ID_comp} could lead to large behavior differences after several orbits. In this section we study the evolution of non-spinning ID sets produced by our method and compare them with those produced with {\it LORENE} library. As previously mentioned, the goal of these simulations is not to provide highly accurate models of the mergers but simply to test the ID sets: the simulations track the BNS evolution until the code crashes due to the formation of singularities and no attempt is made to study in detail the formation/evolution of the newly created compact object and/or detect horizons.

\subsubsection{Varying binary separation}
\label{varying_sep}

We start our tests by evolving sets with coordinate separations of $60$ km ({\it BNS60n}) and $80$ km ({\it BNS80n}) and compared them with their {\it LORENE} counterparts. These simulations' parameters are given in the second block of Table \ref{Table1}.
\begin{figure}[ht]
\centering
\includegraphics[width=0.5\textwidth,keepaspectratio]{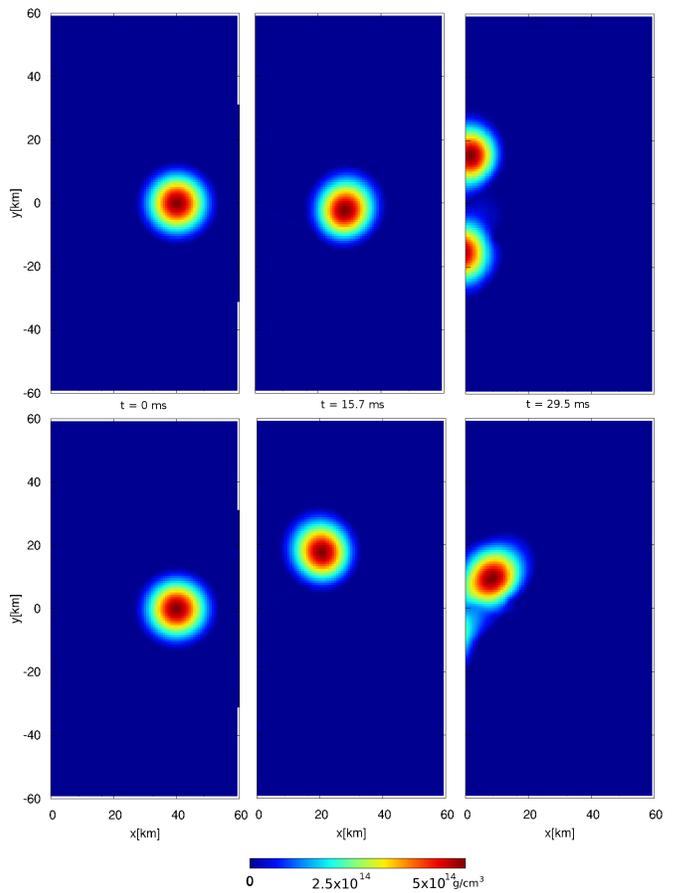}
\caption{Rest mass density for the {\it LORENE80} (top) and the {\it BNS80n} (bottom) simulations.}
\label{rho_xy_plane}
\end{figure}
\begin{figure}[h]
        \includegraphics[width=0.4\textwidth,keepaspectratio]{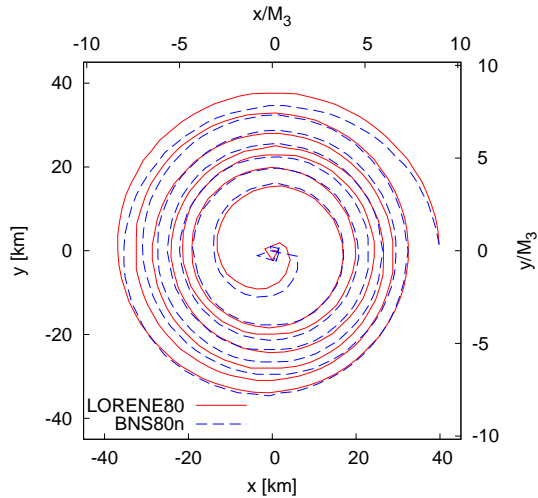}
        \caption{Trajectory of one NS for the runs of Fig. \ref{rho_xy_plane}.
           \label{trajectory_comp}}
\end{figure}
\begin{figure}[h]
        \includegraphics[width=0.5\textwidth,keepaspectratio]{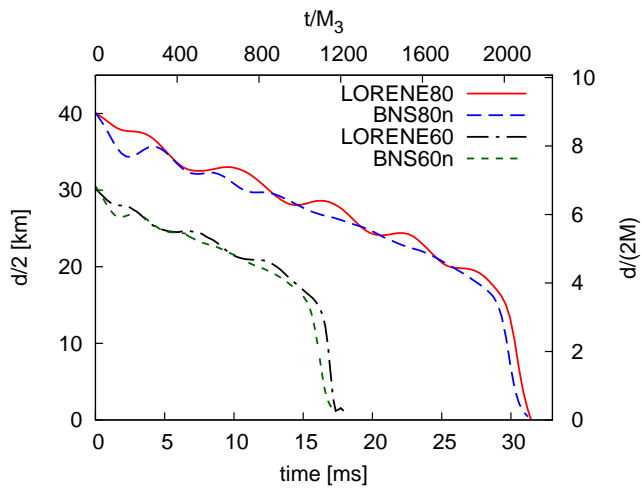}
        \caption{Half of the coordinate separation between NS centers for equal-mass non-spinning binaries. The runs correspond to initial coordinate separations of $60$ and $80$ km.
           \label{sep_comp}}
\end{figure}
\begin{figure}[h]
        \includegraphics[width=0.5\textwidth,keepaspectratio]{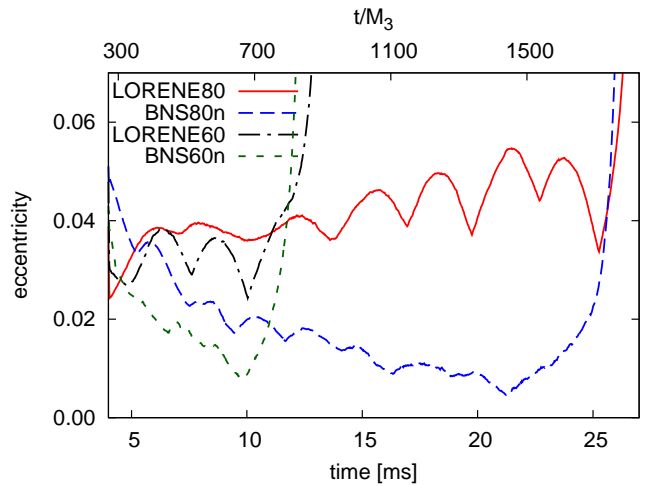}
        \caption{Eccentricity estimation for the runs of Fig. \ref{sep_comp}. The eccentricity was calculated using formula (1) from \cite{Tichy:2010qa}.
\label{ecc_comp}}
\end{figure}
\begin{figure}[h]
    \includegraphics[width=0.5\textwidth,keepaspectratio]{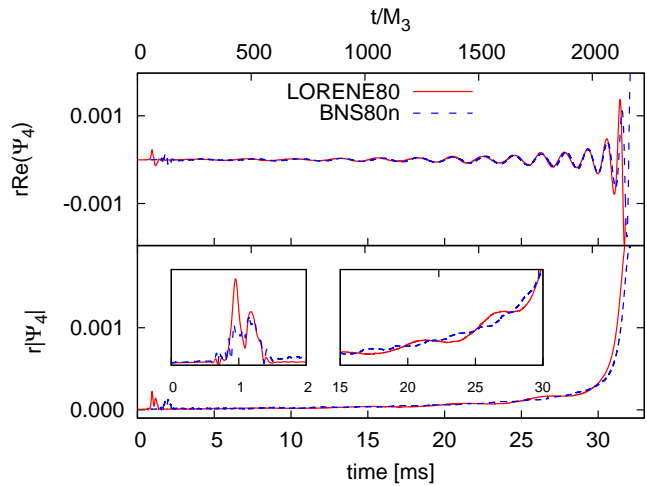}
        \caption{Gravitational wave amplitude (bottom) and real part (top) of  
        the $(2,2)$ mode of $\Psi_4$ for the runs of Fig. \ref{rho_xy_plane}. Signals were shifted to coincide at the amplitude maximum (top and bottom) and at the ``junk" radiation (lower left inset).    
\label{GW_comp}}
\end{figure}
\begin{figure}[h]
        \includegraphics[width=0.5\textwidth]{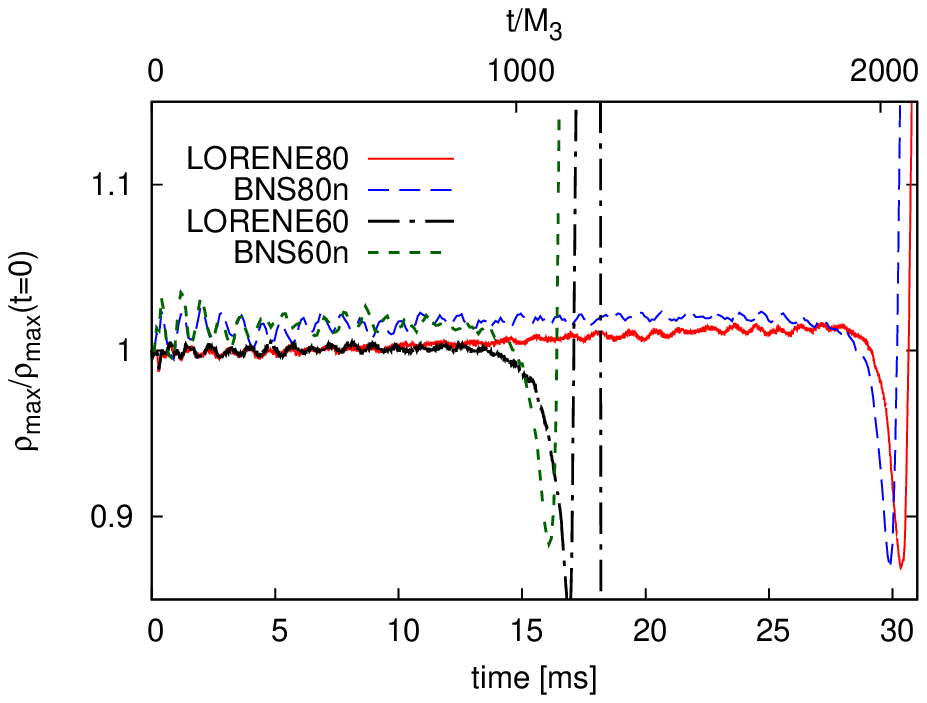}
        \caption{Maximum rest mass density normalized to the initial values for the runs of Fig. \ref{sep_comp}.
\label{rho_comp}}
\end{figure}
\begin{figure}[ht]
\centering
\includegraphics[width=0.5\textwidth,keepaspectratio]{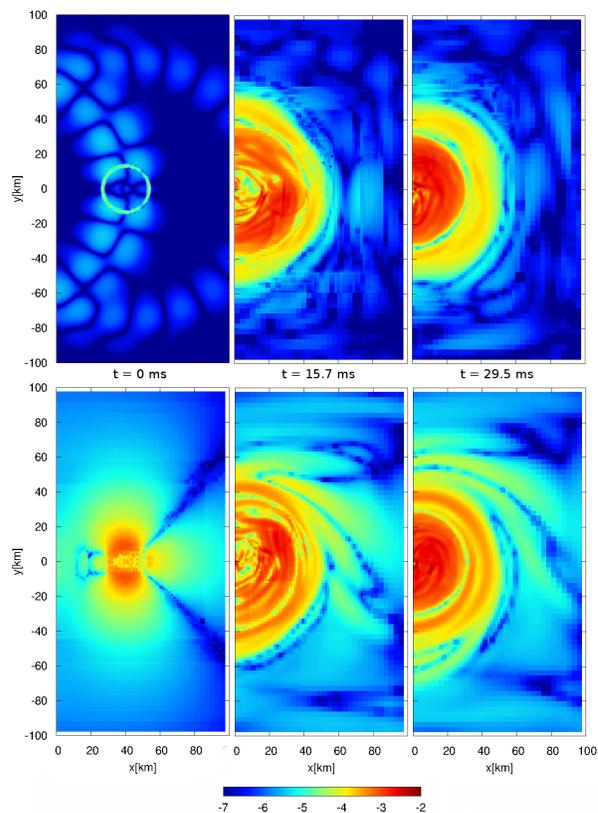}
\caption{Absolute value of the Hamiltonian constraint violation for the {\it Lorene80} (top) and the {\it BNS80n} (bottom) simulations. The time of the snapshots match those of Fig. \ref{rho_xy_plane}. The color logarithmic ruler is shown at the bottom.}
\label{HC_xy_plane_wide}
\end{figure}
\begin{figure}[h]
        \includegraphics[width=0.5\textwidth,keepaspectratio]{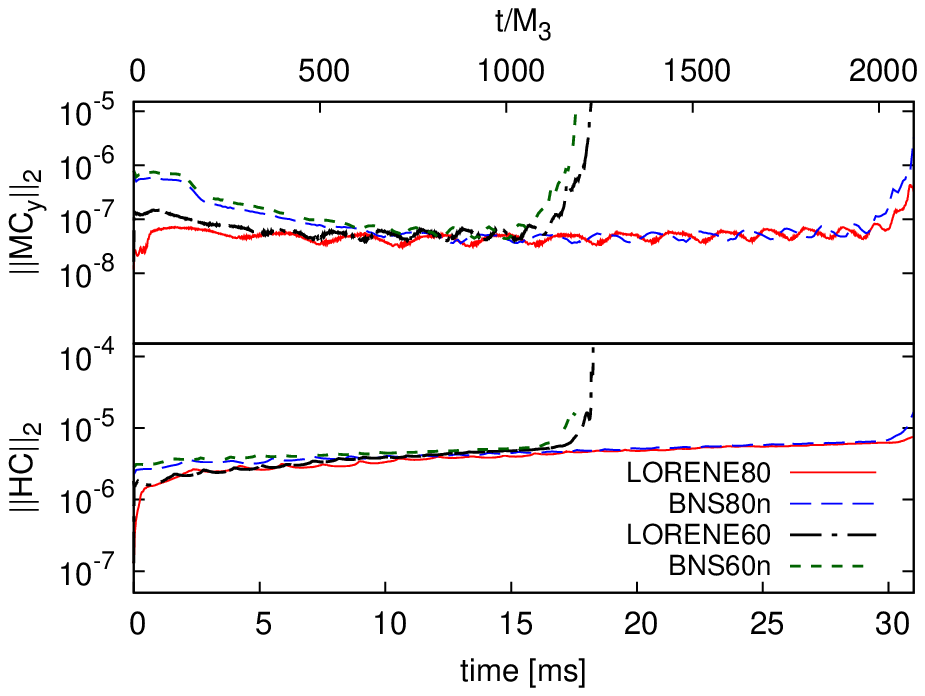}
        \caption{The $L_2$-norm of the Hamiltonian (top) and the
      $y$-component of the momentum (bottom) constraint violations for the runs of Fig. \ref{sep_comp}.   
\label{constr_comp}}
\end{figure}

While some of the design parameters of the {\it BNS} sets were chosen to produce a meaningful comparison with the corresponding {\it LORENE} sets (coordinate separation, total rest mass, zero spins), the boost velocities were fine-tuned to minimize eccentricity. The purpose was to find ID sets less eccentric than those based on helical symmetry. This idea was inspired by the work of Miller \cite{Miller03c} who showed that zeroing the radial velocity could lead to non-negligible orbital eccentricities. More recent work done on BBH \cite{Husa:2007rh,Walther:2009ng,Pfeiffer:2007yz,Boyle:2007ft,Mroue:2010re} and BNS \cite{Kiuchi:2009jt} ID sets show that this eccentricity can be controlled to some extent by a careful choice of tangential and transverse velocities. The determination of these velocity components was done by a corrective iteration that started with a null radial velocity $v^r$ and a post-Newtonian estimation \cite{Kidder:1995zr} for the tangential velocity $v^\phi$ \footnote{Numerical experiments have shown that, for the grids employed here, the best initial guess for the tangential component is $15\%$ larger than the post$^{5/2}$-{N}ewtonian estimation.}. The iteration alternates adjustments for both components using a bisection-like algorithm that, in average, bracketed the velocity components after ten iterations. The boost velocity determination could be sped up by adopting more efficient techniques such as the methods developed for BBH ID sets \cite{Pfeiffer:2007yz,Tichy:2010qa} and this will be explored in future work. The optimal boost velocity for each {\it BNS} set evolved for this article is given in Table \ref{Table1}. 

Figure \ref{rho_xy_plane} shows three snapshots of the rest mass density for the {\it BNS80n} and the {\it LORENE80} simulations, while Fig. \ref{trajectory_comp} plots the trajectory of one of the NS for each simulation. Figure \ref{sep_comp} compares the evolution of half of the coordinate separation between NS centers for the {\it BNS60n} and {\it BNS80n} runs and the respective {\it LORENE} counterparts. We see that the orbital eccentricities of the {\it BNS} runs are smaller than those of the {\it LORENE} runs; a difference that can be quantitatively appreciated in the eccentricity plot of Fig. \ref{ecc_comp} (the eccentricity was calculated using the formula ($1$) from \cite{Tichy:2010qa}). By the time of the last orbits, the {\it BNS} runs' eccentricities are close to an order of magnitude smaller than those of the {\it LORENE} simulations. Note that the method of adjusting the orbital velocity to reduce eccentricity can also be applied to {\it LORENE}-like ID sets, as it has been shown by Kiuchi \et \cite{Kiuchi:2009jt}. Note that in these cases, the Hamiltonian and momentum constraint equations have to be re-solved after the modification is applied. 

The coordinate systems used in the {\it LORENE} and {\it BNS} ID sets are essentially different since they depend on the choice of lapse and shift: while $\alpha$ and $\beta^i$ are freely specifiable when using superposition, they are obtained by numerically solving coupled elliptic equations in the {\it LORENE} ID sets. In order to confirm that the eccentricity reduction is not a coordinate effect, we computed the gravitational waves resulting from the binary evolution at an extraction coordinate radius of $200M_\odot$. Figure \ref{GW_comp} shows the corresponding GW where, for clarity, the signals were shifted in time to make the maxima in amplitude coincide. The eccentricity reduction is appreciated in the amplitude inset in the bottom panel of the figure. Additionally, our ID sets seem to possess less ``junk" radiation. Whether this effect persists in the case of BNS with generic spins will be studied in future work.

To further assess the quality of the {\it BNS} simulations, we plotted the maximum rest mass density as a function of time (Fig. \ref{rho_comp}). Our ID runs show oscillations with larger amplitude than those of {\it LORENE} during the first orbits; a behavior already discussed in Fig. \ref{over_contraction}. However, the amplitude diminishes with time and, after a couple of orbits all the runs show comparable oscillations. The most important question for the assessment of the quality of our ID sets is, however, how large are the violations of the constraints? Figure \ref{HC_xy_plane_wide} presents a qualitative view of the Hamiltonian constraint violation through snapshots that compare {\it LORENE80} vs. {\it BNS80n}. A quantitative comparison is given by the $L_2$ norm of the Hamiltonian and the $y$ component of the momentum constraint violations as a function of time (Fig. \ref{constr_comp}). As expected, at $t=0$ the constraint violations in the {\it BNS} runs are larger than those of the {\it LORENE} counterparts. However, after a couple of orbits, the difference diminishes and the scale of the violations for all the runs become comparable. Other quantitative indicators such as the $L_1$ and $L_\infty$ norms show similar behavior. This  damping is attributable to the well known constraint violation propagation properties of the BSSNOK formulation and seems to indicate that this violation reduction could possibly be hasten with the use of formulations with superior constraint damping characteristics such as CCZ4 \cite{Bernuzzi:2009ex, Alic:2011gg} and Z4c \cite{Hilditch:2012fp} (Kastaun \et \cite{Kastaun:2013mv} and Alic \et \cite{Alic:2013xsa} show CCZ4 simulations where the constraint violations fall well below BSSNOK levels after only 1ms).

\begin{figure}[h]
        \includegraphics[width=0.5\textwidth,keepaspectratio]{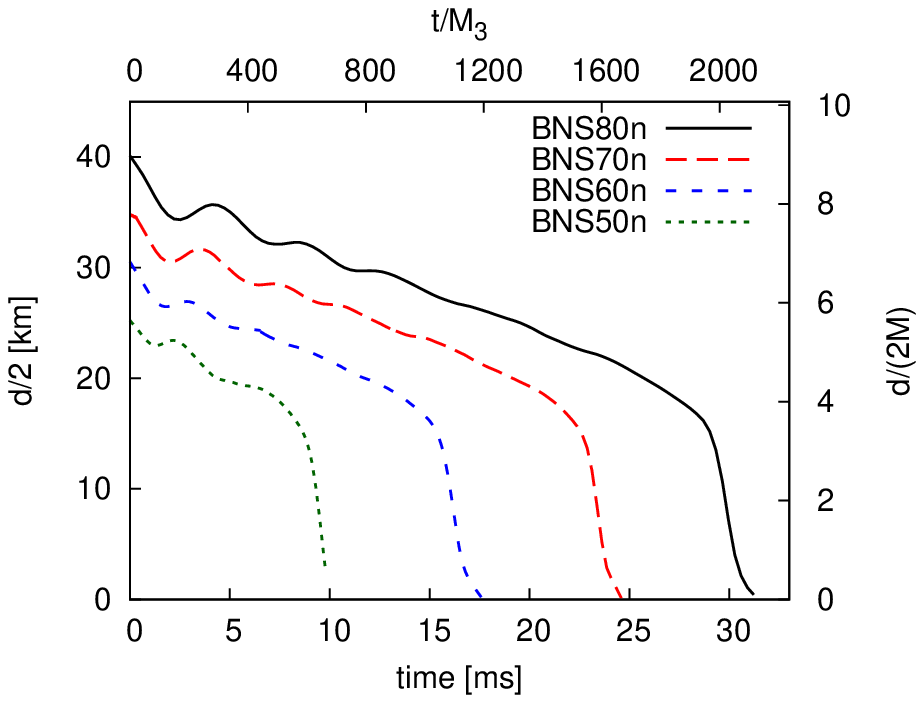}
        \caption{Comparison of {\it BNS} ID evolutions starting at different coordinate separations. These runs are described in the second block of Table \ref{Table1}.}
\label{BNS_sep_study_sep}
\end{figure}
\begin{figure}[h]
        \includegraphics[width=0.5\textwidth,keepaspectratio]{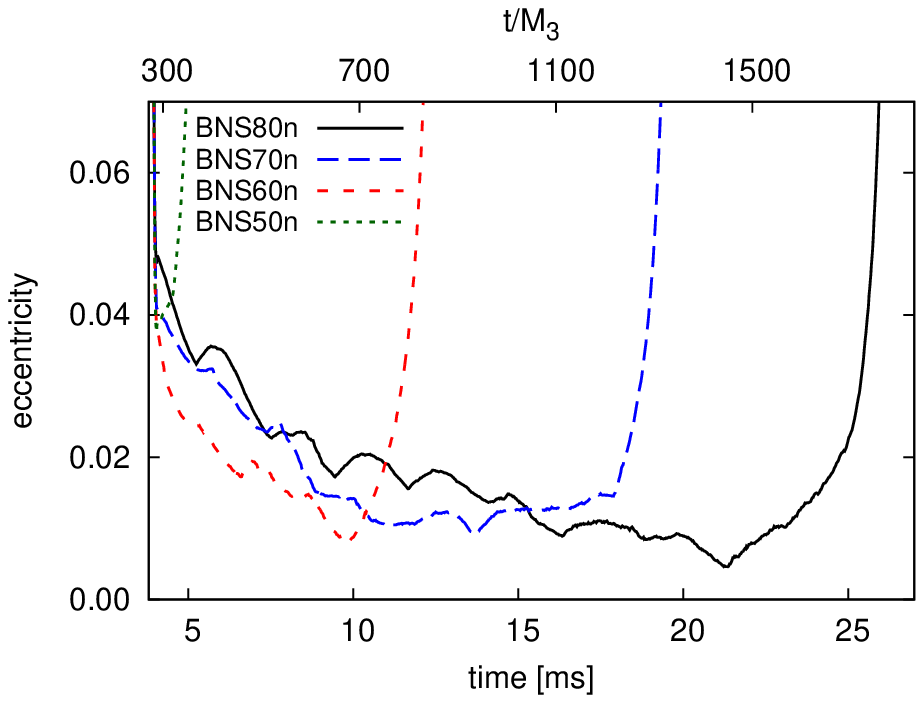}
        \caption{Eccentricity estimation for the runs of Fig. \ref{BNS_sep_study_sep}.}
\label{BNS_sep_study_ecc}
\end{figure}
Finally, we added to this section a series of evolutions of {\it BNS} ID sets corresponding to non-spinning NS starting at separation distances between $50$ and $80$ km. Figures \ref{BNS_sep_study_sep} and \ref{BNS_sep_study_ecc} show the coordinate separation and eccentricity as a function of time. Details of these runs are provided in the second block of Table \ref{Table1}. The evolution of the run starting at a coordinate separation of $50$ km is too short to provide a meaningful measure of the eccentricity. Note that, against expectations, the absolute value of the boost velocity radial component seems to increase with the separation. We attribute this to the dominance of grid structure effects, in particular resolution, for the setups employed here.

\subsubsection{Varying grid resolution}
\label{varying_res}

A complete study of how these quality control markers depend on the grid resolution is important and, due to the corresponding large computational cost, outside the scope of this work. However, to gain insights on the behavior of simulations based on our data at resolutions similar to those of the current state-of-the-art simulations, we have performed a set of runs corresponding to equal mass non-spinning binaries that start at a separation of $60$ km. These runs are listed in the third block of Table \ref{Table1}, where the ``lr", ``mr" and ``hr" suffixes indicate low ($\Delta x=0.50M_{\odot}$), medium ($\Delta x=0.25M_{\odot}$) and high ($\Delta x=0.1875M_{\odot}$) maximum spatial resolution. The grid structure is similar in this cases with the addition of an extra level of refinement at the center of the medium and high resolution runs.

\begin{figure}[h]
        \includegraphics[width=0.5\textwidth,keepaspectratio]{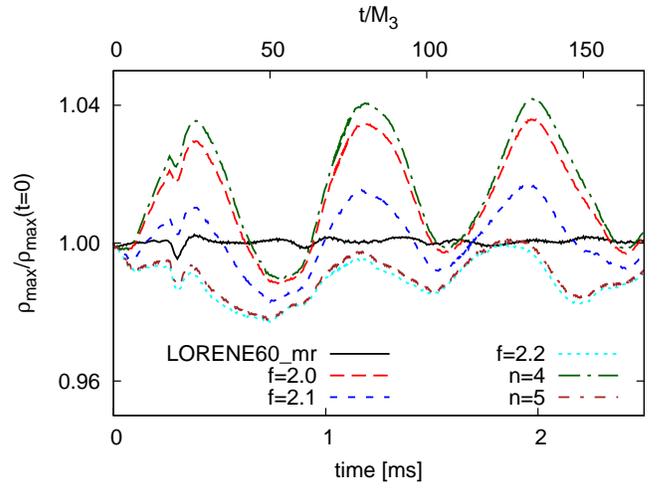}
        \caption{Comparison between the over-contraction method of Eq (\ref{n_LT}), represented by the curves with the parameter $n$, and that of Eq. (\ref{f_LT}), represented by the curves with the parameter $f$. All the {\it BNS} simulations were performed with the same parameters used for {\it BNS60n\_mr}, with the obvious exception of the values of $n$ and $f$.}
\label{Comp_n_ocf}
\end{figure}
There is an additional difference between the runs in this section and the previous ones: these ID sets are based on the alternative scheme of Eq. (\ref{f_LT}) instead of that of Eq. (\ref{n_LT}). The advantage of using Eq. (\ref{f_LT}) over Eq. (\ref{n_LT}) can be better appreciated in the medium resolution runs displayed in Fig. \ref{Comp_n_ocf}. There we see that the oscillations of the central density observed when using $n=4$ ($n=5$) in Eq. (\ref{n_LT}) are similar to those obtained when using $f=2.0$ ($f=2.2$) in Eq. (\ref{f_LT}). A value of $f=2.1$ provides intermediate amplitude variations of the order of $1\%$ and is the value selected for the {\it BNS60n\_lr}, {\it BNS60n\_mr} and {\it BNS60n\_hr} runs.
\begin{figure}[h]
        \includegraphics[width=0.5\textwidth,keepaspectratio]{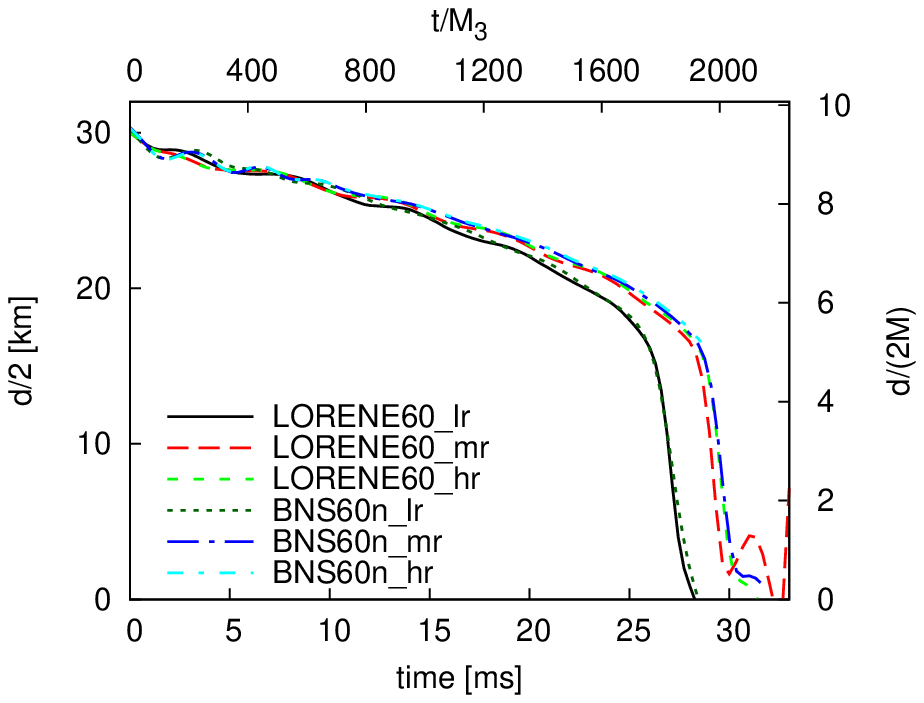}
        \caption{Half of the coordinate separation between NS centers for equal-mass non-spinning binaries. The runs correspond to binaries with initial separation of $60$ km and compare low, medium and high resolution results.
\label{sep_comp2}}
\end{figure}
\begin{figure}[h]
        \includegraphics[width=0.5\textwidth,keepaspectratio]{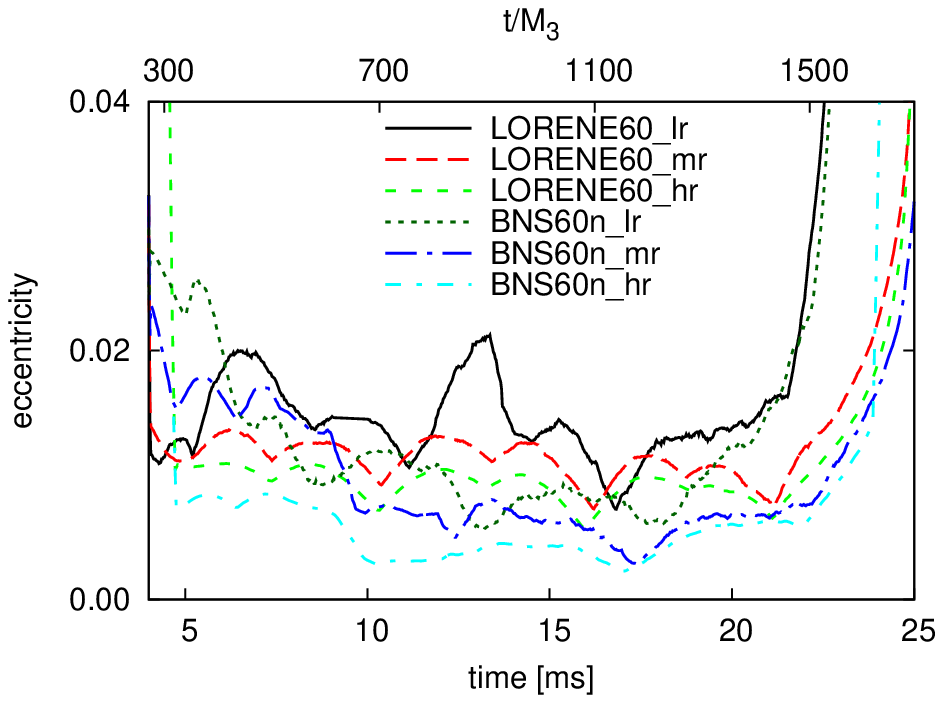}
        \caption{Eccentricity estimation for the runs of Fig. \ref{sep_comp2}.
\label{ecc_comp2}}
\end{figure}
\begin{figure}[h]
        \includegraphics[width=0.5\textwidth]{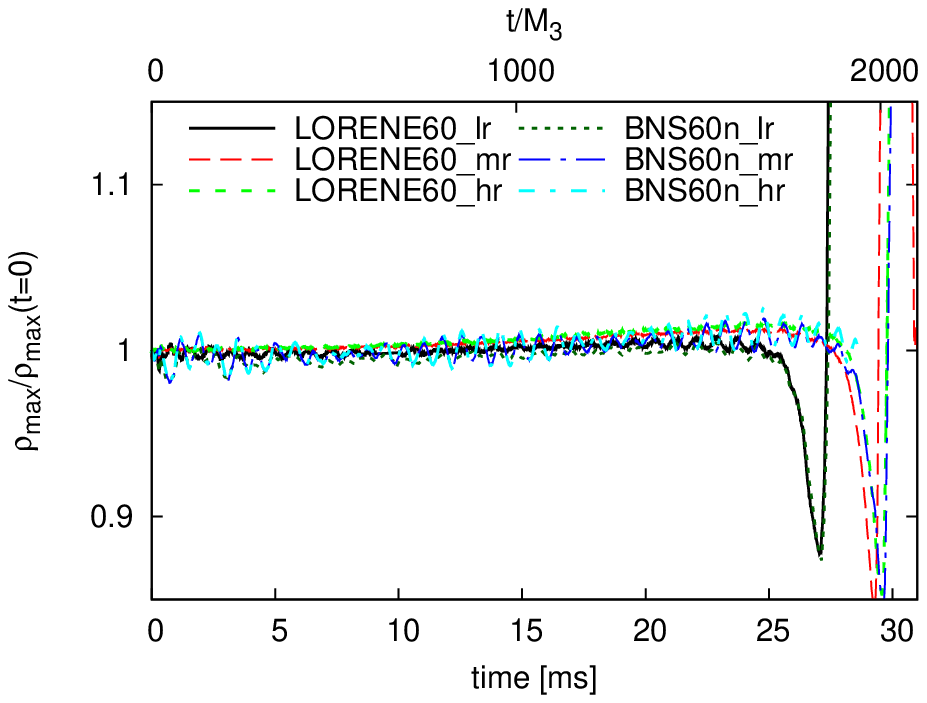}
        \caption{Maximum rest mass density normalized to the initial values for the runs of Fig. \ref{sep_comp2}.
\label{rho_comp2}}
\end{figure}
\begin{figure}[h]
        \includegraphics[width=0.5\textwidth,keepaspectratio]{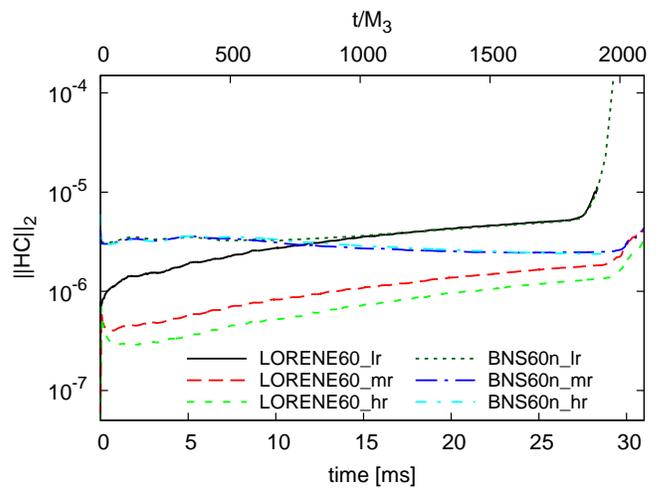}
        \caption{The $L_2$-norm of the Hamiltonian constraint violation for the runs of Fig. \ref{sep_comp2}.   
\label{constr_comp2}}
\end{figure}
\begin{figure}[h]
        \includegraphics[width=0.5\textwidth,keepaspectratio]{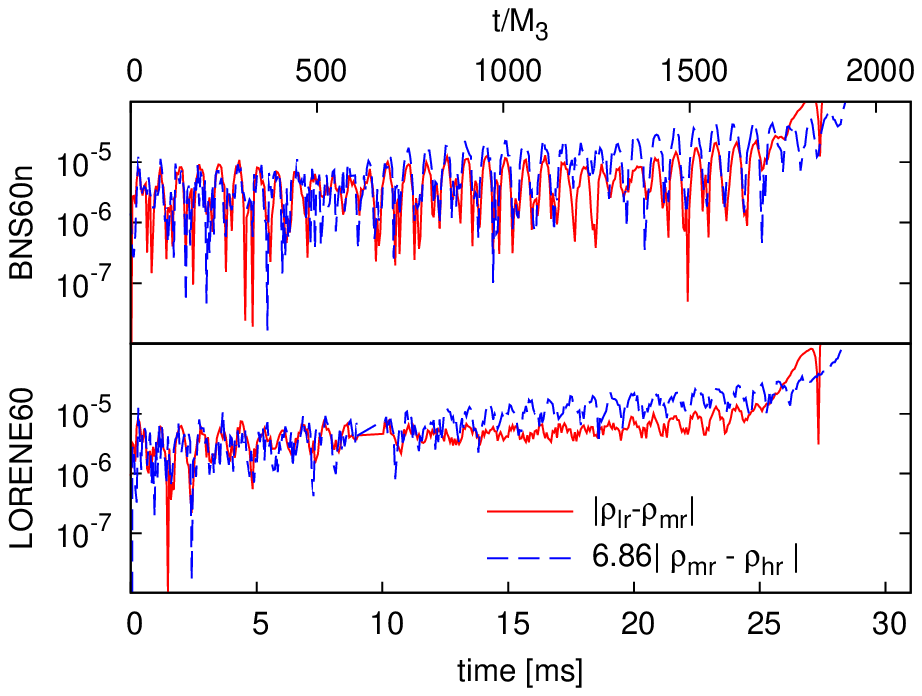}
        \caption{Study of the convergence of the maximum rest mass density with resolution for the runs of Fig. \ref{sep_comp2}. The factor multiplying the dashed curves corresponds to that of second order convergence.
\label{dens_conv}}
\end{figure}

Figures \ref{sep_comp2} and \ref{ecc_comp2} show the evolution of the coordinate separation and eccentricity for the runs described in the third block of Table \ref{Table1}. These six simulations were run using gauge ($\alpha_d, ~\beta_d$) and dissipation ($\epsilon_{diss}$) parameters that resemble more closely those employed in the most current binary modeling performed with ET. One clear effect of this set of parameters is the reduction of eccentricity in the {\it LORENE60} runs. Our choice for the boost velocity used for the corresponding {\it BNS60n} runs manages to reduce the orbital eccentricity even further, but this reduction is not as large as the one presented in the previous section. It can be seen, however, that increasing the spatial resolution diminishes the orbital eccentricity for both types of ID. In the {\it BNS60n} ID cases, smaller eccentricities could potentially be achieved by further fine-tuning of the boost velocity. Again we relied on comparisons of the evolution of the maximum rest mass density (Fig. \ref{rho_comp2}) and the constraints violations (Fig. \ref{constr_comp2}) to assess the behavior of our ID sets under time evolutions. The results are qualitatively similar to those of the previous section. The main difference is that the time needed for the Hamiltonian constraint violations of the {\it BNS60n} runs to relax to the level of the corresponding {\it LORENE60} cases increases with decreasing resolution. This is to be expected: the use of higher grid resolution diminishes the initial constraint violation of the {\it LORENE60} ID sets (which are, after all, numerical solutions of the constraints) while leaving mostly intact that of the {\it BNS60n} cases (only an approximation). As mentioned above, the use of formulations such as CCZ4 and Z4c has been shown to reduce this relaxation time to the point where this difference may be moot for simulations at the resolution used in the current state-of-the-art runs \cite{Kastaun:2013mv, Alic:2013xsa}. Finally, Fig. \ref{dens_conv} shows that the evolution of the maximum rest mass density is consistent with a second order convergence in resolution: while the spatial stencils used here are fourth order, hydrodynamics of compact objects usually present a degree of degradation due to stellar surface effects.

\subsection{Spinning binaries evolution}
\label{spinning_BNS}

\begin{figure}
 \includegraphics[width=0.45\textwidth,keepaspectratio]{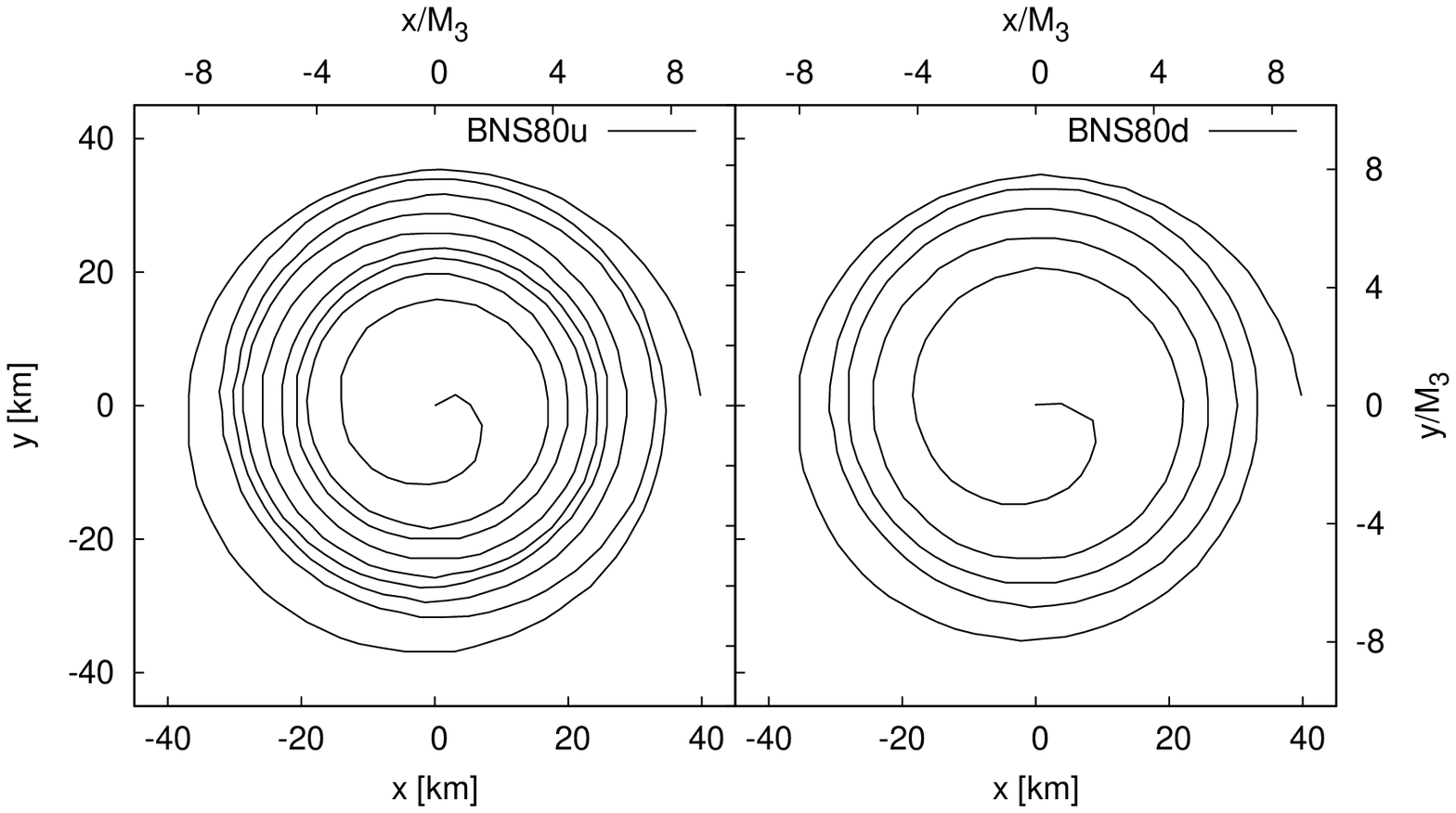}
 \caption{Trajectory of one NS for the evolution of binary sets with spins aligned ({\it BNS80u}, left) and anti-aligned ({\it BNS80d}, right) with the orbital angular momentum that start at a coordinate separation of $80$ km.} 
\label{trajectory_spinning}
\end{figure}
\begin{figure}[h]
   \includegraphics[width=0.5\textwidth,keepaspectratio]{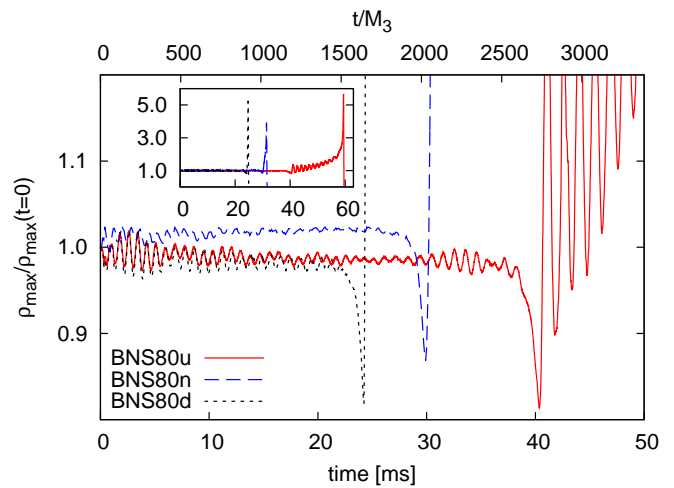}
   \caption{Maximum rest mass density normalized to the initial value for simulations of binaries with spins aligned ({\it BNS80u}), zero ({\it BNS80n}) and anti-aligned ({\it BNS80d}) with the orbital angular momentum.    
    \label{rho_spin}}
\end{figure} 
\begin{figure}[h]
    \includegraphics[width=0.5\textwidth,keepaspectratio]{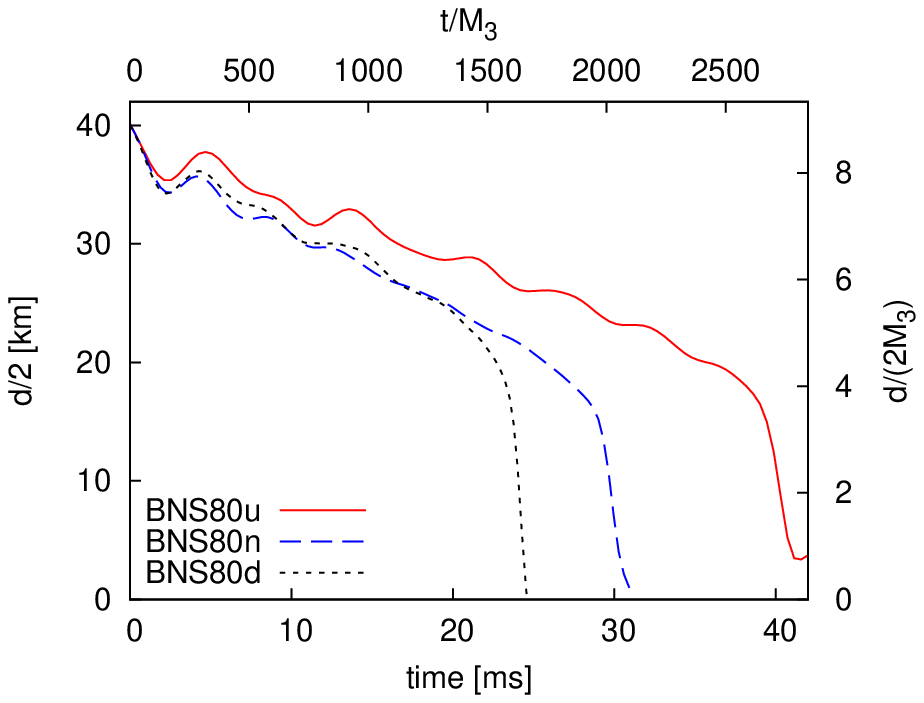}
        \caption{Coordinate separation for the runs of Fig. \ref{rho_spin}.      
\label{sep_spin}}
\end{figure}
\begin{figure}[h]
    \includegraphics[width=0.5\textwidth,keepaspectratio]{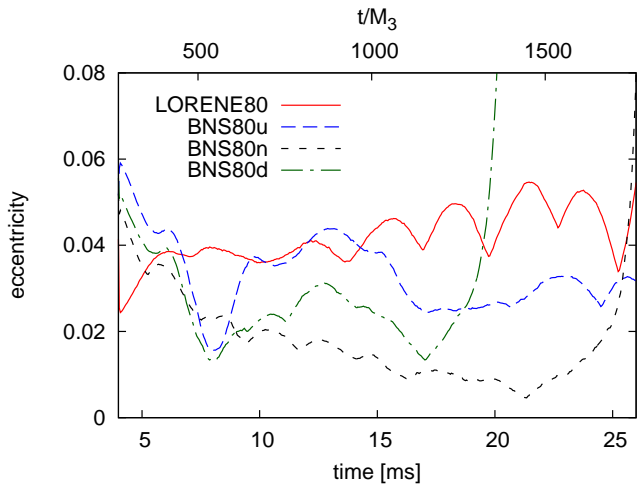}
        \caption{Orbital eccentricity for the runs of Fig. \ref{rho_spin}.      
\label{ecc_spin}}
\end{figure}
\begin{figure}[h]
    \includegraphics[width=0.5\textwidth,keepaspectratio]{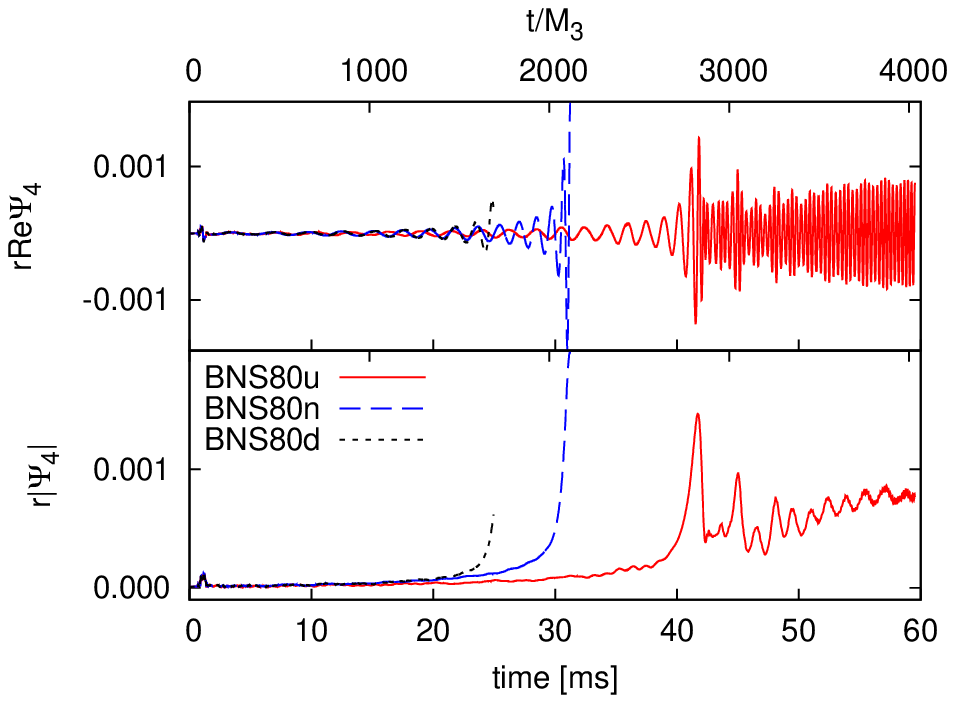}
        \caption{Gravitational wave amplitude (bottom) and real part (top) of  
        the $(2,2)$ mode of $\Psi_4$ for the runs of Fig. \ref{rho_spin}.     
\label{GW_spin}}
\end{figure}

The most salient characteristic of our method for constructing BNS ID sets is the ability to handle spinning stars. To show this, we produced two ID sets with NS with spins aligned ({\it BNS80u}) and anti-aligned ({\it BNS80d}) with the orbital angular momentum and evolved them from a starting coordinate separation of $80$ km through their mergers. Both sets have identical NS with rest masses $M_{0}=1.625M_{\odot}$ and dimensionless spin parameters $a=J/M^2\simeq0.33$, as detailed in last block of Table \ref{Table1}.

Figure \ref{trajectory_spinning} shows the trajectories of one NS for the evolution of each set; these curves complement the non-spinning BNS ({\it BNS80n}) trajectory given in Fig. \ref{trajectory_comp}. One interesting feature is the presence of the orbital ``hang-up" \cite{Campanelli:2006uy}. This effect predicts that systems with spins aligned with the orbital angular momentum orbit longer than those with anti-aligned spins, allowing the shedding of excess of angular momentum through the emission of gravitational radiation. While the anti-aligned binary merged and immediately collapsed to a black hole, the aligned case led to the formation of a centrifugally supported hypermassive neutron star that survived for about $20$ ms before collapsing.

Figures \ref{rho_spin} and \ref{sep_spin} show the maximum rest mass density and the coordinate separation vs. time for the three {\it BNS80} runs. We have made an effort to find the velocities that would minimize the eccentricity but halted the fine-tuning after reaching what we deemed a reasonable precision considering the low resolution of these simulations. More accurate evolutions with even lower eccentricities are likely to result in merger times that are quantitatively different than the values seen here. Figure \ref{ecc_spin} presents the orbital eccentricities of the three {\it BNS80} runs and compares them with that of the {\it LORENE80} simulation. The eccentricities achieved during the last orbits of the three {\it BNS80} cases are lower than that of the reference {\it LORENE80} simulation.

\section{Conclusions}
\label{conclusions}

We introduced a new way of constructing initial data for binary neutron stars with arbitrary spins and orbital eccentricities. The method only offers approximations to the Einstein field equations since, by design, the data sets do not satisfy the Hamiltonian and momentum constraints. However, by evolving our initial data using the BSSNOK formulation, we showed that these constraint violations become comparable to those seen in evolutions of standard (i.e., irrotational, conformally-flat, helically symmetric, constraint solving) initial data sets after relaxation times that increase with increasing grid resolution. Our method consists of a variant of metric superposition that addresses two common problems: large stellar shape oscillations and orbital eccentricities. It reduces the former to variations of the order of $1\%$ and offers great control over orbital eccentricities. Additionally, we see indications that our initial data sets possess less ``junk" radiation than that found in standard sets. 

We tested our initial data by evolving in time initial data for single and binary neutron star systems. We showed that our data leads to inspirals with orbital eccentricities smaller than those seen in standard initial data simulations. However, since the method's most important characteristic is the ability to handle spinning binaries, we also evolved binaries with spins aligned and anti-aligned with the orbital angular momentum. The anti-aligned binary merges and immediately collapses to a black hole, while the aligned case leads to the formation of a centrifugally supported hypermassive neutron star that survives for several dynamical times before collapsing. 

The work presented here will be followed by studies that are outside of the scope of this article due to computational demands. We plan to explore the viability of our method for binaries with generic spins and explore possible improvements aimed at reducing the constraint violations and orbital eccentricity even further. We will study more efficient ways of selecting the stars' boost velocity since the direct trial-and-error approach employed here is time and resource consuming. We will conduct a more systematical study of the content of ``junk" radiation in our sets, to find out if the reduction in this unwanted quantity is a common feature in binaries with arbitrary spins. Another priority is the evolution of our ID sets using numerical formulations such as CCZ4 and Z4c, since they possess better contraint damping characteristics than BSSNOK. Finally, more realistic simulations (different equations of state, horizon detection and tracking, ringdown modeling) with arbitrarily spinning stars (which would require grids without symmetries) will be pursued.

\begin{acknowledgments}
It is a pleasure to thank K. Yakunin for useful discussions. We are also grateful to S. Bernuzzi, T. Dietrich, F. Galeazzi, T. Font, J. Friedman, W. Tichy and in particular L. Rezzolla for comments on the manuscript. This research was supported by the NSF award PHY-0855315 and by an allocation of advanced computing resources provided by the National Science Foundation. The computations were performed on Kraken at the National Institute for Computational Sciences (http://www.nics.tennessee.edu/).
\end{acknowledgments}

\pagebreak[4]

\bibliography{references}

\begin{thebibliography}{68}
\expandafter\ifx\csname natexlab\endcsname\relax\def\natexlab#1{#1}\fi
\expandafter\ifx\csname bibnamefont\endcsname\relax
  \def\bibnamefont#1{#1}\fi
\expandafter\ifx\csname bibfnamefont\endcsname\relax
  \def\bibfnamefont#1{#1}\fi
\expandafter\ifx\csname citenamefont\endcsname\relax
  \def\citenamefont#1{#1}\fi
\expandafter\ifx\csname url\endcsname\relax
  \def\url#1{\texttt{#1}}\fi
\expandafter\ifx\csname urlprefix\endcsname\relax\def\urlprefix{URL }\fi
\providecommand{\bibinfo}[2]{#2}
\providecommand{\eprint}[2][]{\url{#2}}

\bibitem[{\citenamefont{Gehrels and M{\'e}sz{\'a}ros}(2012)}]{Gehrels24082012}
\bibinfo{author}{\bibfnamefont{N.}~\bibnamefont{Gehrels}} \bibnamefont{and}
  \bibinfo{author}{\bibfnamefont{P.}~\bibnamefont{M{\'e}sz{\'a}ros}},
  \bibinfo{journal}{Science} \textbf{\bibinfo{volume}{337}},
  \bibinfo{pages}{932} (\bibinfo{year}{2012}).

\bibitem[{\citenamefont{Gomboc}(2012)}]{Gomboc:2012ff}
\bibinfo{author}{\bibfnamefont{A.}~\bibnamefont{Gomboc}}
  (\bibinfo{year}{2012}), \eprint{1206.3127}.

\bibitem[{\citenamefont{Sathyaprakash and Schutz}(2009)}]{Sathyaprakash:2009xs}
\bibinfo{author}{\bibfnamefont{B.}~\bibnamefont{Sathyaprakash}}
  \bibnamefont{and} \bibinfo{author}{\bibfnamefont{B.}~\bibnamefont{Schutz}},
  \bibinfo{journal}{Living Rev.Rel.} \textbf{\bibinfo{volume}{12}},
  \bibinfo{pages}{2} (\bibinfo{year}{2009}), \eprint{0903.0338}.

\bibitem[{\citenamefont{Abadie et~al.}(2010)}]{Abadie:2010cf}
\bibinfo{author}{\bibfnamefont{J.}~\bibnamefont{Abadie}} \bibnamefont{et~al.}
  (\bibinfo{collaboration}{LIGO Scientific Collaboration, Virgo
  Collaboration}), \bibinfo{journal}{Class.Quant.Grav.}
  \textbf{\bibinfo{volume}{27}}, \bibinfo{pages}{173001}
  (\bibinfo{year}{2010}), \eprint{astro-ph/1003.2480}.

\bibitem[{\citenamefont{Wilson and Mathews}(1995)}]{Wilson95}
\bibinfo{author}{\bibfnamefont{J.~R.} \bibnamefont{Wilson}} \bibnamefont{and}
  \bibinfo{author}{\bibfnamefont{G.~J.} \bibnamefont{Mathews}},
  \bibinfo{journal}{Phys. Rev. Lett.} \textbf{\bibinfo{volume}{75}},
  \bibinfo{pages}{4161} (\bibinfo{year}{1995}).

\bibitem[{\citenamefont{Wilson et~al.}(1996)\citenamefont{Wilson, Mathews, and
  Marronetti}}]{Wilson:1996ty}
\bibinfo{author}{\bibfnamefont{J.~R.} \bibnamefont{Wilson}},
  \bibinfo{author}{\bibfnamefont{G.~J.} \bibnamefont{Mathews}},
  \bibnamefont{and}
  \bibinfo{author}{\bibfnamefont{P.}~\bibnamefont{Marronetti}},
  \bibinfo{journal}{Phys. Rev.} \textbf{\bibinfo{volume}{D54}},
  \bibinfo{pages}{1317} (\bibinfo{year}{1996}), \eprint{gr-qc/9601017}.

\bibitem[{\citenamefont{Peters and Mathews}(1963)}]{Peters:1963ux}
\bibinfo{author}{\bibfnamefont{P.~C.} \bibnamefont{Peters}} \bibnamefont{and}
  \bibinfo{author}{\bibfnamefont{J.}~\bibnamefont{Mathews}},
  \bibinfo{journal}{Phys. Rev.} \textbf{\bibinfo{volume}{131}},
  \bibinfo{pages}{435} (\bibinfo{year}{1963}).

\bibitem[{\citenamefont{Oslowski et~al.}(2009)\citenamefont{Oslowski, Bulik,
  Gondek-Rosinska, and Belczynski}}]{Oslowski:2009zr}
\bibinfo{author}{\bibfnamefont{S.}~\bibnamefont{Oslowski}},
  \bibinfo{author}{\bibfnamefont{T.}~\bibnamefont{Bulik}},
  \bibinfo{author}{\bibfnamefont{D.}~\bibnamefont{Gondek-Rosinska}},
  \bibnamefont{and}
  \bibinfo{author}{\bibfnamefont{K.}~\bibnamefont{Belczynski}}
  (\bibinfo{year}{2009}), \eprint{0903.3538}.

\bibitem[{\citenamefont{Kowalska et~al.}(2011)\citenamefont{Kowalska, Bulik,
  Belczynski, Dominik, and Gondek-Rosinska}}]{Kowalska:2010qg}
\bibinfo{author}{\bibfnamefont{I.}~\bibnamefont{Kowalska}},
  \bibinfo{author}{\bibfnamefont{T.}~\bibnamefont{Bulik}},
  \bibinfo{author}{\bibfnamefont{K.}~\bibnamefont{Belczynski}},
  \bibinfo{author}{\bibfnamefont{M.}~\bibnamefont{Dominik}}, \bibnamefont{and}
  \bibinfo{author}{\bibfnamefont{D.}~\bibnamefont{Gondek-Rosinska}},
  \bibinfo{journal}{Astron.Astrophys.} \textbf{\bibinfo{volume}{527}},
  \bibinfo{pages}{A70} (\bibinfo{year}{2011}), \eprint{1010.0511}.

\bibitem[{\citenamefont{Gold et~al.}(2012)\citenamefont{Gold, Bernuzzi,
  Thierfelder, Brugmann, and Pretorius}}]{Gold:2011df}
\bibinfo{author}{\bibfnamefont{R.}~\bibnamefont{Gold}},
  \bibinfo{author}{\bibfnamefont{S.}~\bibnamefont{Bernuzzi}},
  \bibinfo{author}{\bibfnamefont{M.}~\bibnamefont{Thierfelder}},
  \bibinfo{author}{\bibfnamefont{B.}~\bibnamefont{Brugmann}}, \bibnamefont{and}
  \bibinfo{author}{\bibfnamefont{F.}~\bibnamefont{Pretorius}},
  \bibinfo{journal}{Phys.Rev.} \textbf{\bibinfo{volume}{D86}},
  \bibinfo{pages}{121501} (\bibinfo{year}{2012}), \eprint{1109.5128}.

\bibitem[{\citenamefont{Miller}(2004)}]{Miller03c}
\bibinfo{author}{\bibfnamefont{M.}~\bibnamefont{Miller}},
  \bibinfo{journal}{Phys. Rev. D} \textbf{\bibinfo{volume}{69}},
  \bibinfo{pages}{124013} (\bibinfo{year}{2004}),
  \bibinfo{note}{gr-qc/0305024}.

\bibitem[{\citenamefont{Baumgarte et~al.}(1997)\citenamefont{Baumgarte, Cook,
  Scheel, Shapiro, and Teukolsky}}]{Baumgarte:1997xi}
\bibinfo{author}{\bibfnamefont{T.~W.} \bibnamefont{Baumgarte}},
  \bibinfo{author}{\bibfnamefont{G.~B.} \bibnamefont{Cook}},
  \bibinfo{author}{\bibfnamefont{M.~A.} \bibnamefont{Scheel}},
  \bibinfo{author}{\bibfnamefont{S.~L.} \bibnamefont{Shapiro}},
  \bibnamefont{and} \bibinfo{author}{\bibfnamefont{S.~A.}
  \bibnamefont{Teukolsky}}, \bibinfo{journal}{Phys. Rev. Lett.}
  \textbf{\bibinfo{volume}{79}}, \bibinfo{pages}{1182} (\bibinfo{year}{1997}),
  \eprint{gr-qc/9704024}.

\bibitem[{\citenamefont{Baumgarte et~al.}(1998)\citenamefont{Baumgarte, Cook,
  Scheel, Shapiro, and {T}eukolsky}}]{Baumgarte98c}
\bibinfo{author}{\bibfnamefont{T.~W.} \bibnamefont{Baumgarte}},
  \bibinfo{author}{\bibfnamefont{G.~B.} \bibnamefont{Cook}},
  \bibinfo{author}{\bibfnamefont{M.~A.} \bibnamefont{Scheel}},
  \bibinfo{author}{\bibfnamefont{S.~L.} \bibnamefont{Shapiro}},
  \bibnamefont{and} \bibinfo{author}{\bibfnamefont{S.~A.}
  \bibnamefont{{T}eukolsky}}, \bibinfo{journal}{Phys. Rev. D}
  \textbf{\bibinfo{volume}{57}}, \bibinfo{pages}{6181} (\bibinfo{year}{1998}).

\bibitem[{\citenamefont{Bildsten and Cutler}(1992)}]{Bildsten92}
\bibinfo{author}{\bibfnamefont{L.}~\bibnamefont{Bildsten}} \bibnamefont{and}
  \bibinfo{author}{\bibfnamefont{C.}~\bibnamefont{Cutler}},
  \bibinfo{journal}{Astrophys. J.} \textbf{\bibinfo{volume}{400}},
  \bibinfo{pages}{175} (\bibinfo{year}{1992}).

\bibitem[{\citenamefont{Kochanek}(1992)}]{Kochanek92}
\bibinfo{author}{\bibfnamefont{C.}~\bibnamefont{Kochanek}},
  \bibinfo{journal}{Astrophys. J.} \textbf{\bibinfo{volume}{398}},
  \bibinfo{pages}{234} (\bibinfo{year}{1992}).

\bibitem[{\citenamefont{Bonazzola et~al.}(1997)\citenamefont{Bonazzola,
  Gourgoulhon, and Marck}}]{Bonazzola97}
\bibinfo{author}{\bibfnamefont{S.}~\bibnamefont{Bonazzola}},
  \bibinfo{author}{\bibfnamefont{E.}~\bibnamefont{Gourgoulhon}},
  \bibnamefont{and} \bibinfo{author}{\bibfnamefont{J.-A.} \bibnamefont{Marck}},
  \bibinfo{journal}{Phys. Rev. D} \textbf{\bibinfo{volume}{56}},
  \bibinfo{pages}{7740} (\bibinfo{year}{1997}).

\bibitem[{\citenamefont{Asada}(1998)}]{PhysRevD.57.7292}
\bibinfo{author}{\bibfnamefont{H.}~\bibnamefont{Asada}},
  \bibinfo{journal}{Phys. Rev. D} \textbf{\bibinfo{volume}{57}},
  \bibinfo{pages}{7292} (\bibinfo{year}{1998}),
  \urlprefix\url{http://link.aps.org/doi/10.1103/PhysRevD.57.7292}.

\bibitem[{\citenamefont{Teukolsky}(1998)}]{Teukolsky98}
\bibinfo{author}{\bibfnamefont{S.}~\bibnamefont{Teukolsky}},
  \bibinfo{journal}{Astrophys. J.} \textbf{\bibinfo{volume}{504}},
  \bibinfo{pages}{442} (\bibinfo{year}{1998}).

\bibitem[{\citenamefont{Shibata}(1998)}]{Shibata98}
\bibinfo{author}{\bibfnamefont{M.}~\bibnamefont{Shibata}},
  \bibinfo{journal}{Phys. Rev. D} \textbf{\bibinfo{volume}{58}},
  \bibinfo{pages}{024012} (\bibinfo{year}{1998}).

\bibitem[{\citenamefont{Marronetti et~al.}(1998)\citenamefont{Marronetti,
  Mathews, and Wilson}}]{Marronetti:1998xv}
\bibinfo{author}{\bibfnamefont{P.}~\bibnamefont{Marronetti}},
  \bibinfo{author}{\bibfnamefont{G.~J.} \bibnamefont{Mathews}},
  \bibnamefont{and} \bibinfo{author}{\bibfnamefont{J.~R.}
  \bibnamefont{Wilson}}, \bibinfo{journal}{Phys. Rev.}
  \textbf{\bibinfo{volume}{D58}}, \bibinfo{pages}{107503}
  (\bibinfo{year}{1998}), \eprint{gr-qc/9803093}.

\bibitem[{\citenamefont{Bonazzola et~al.}(1999)\citenamefont{Bonazzola,
  Gourgoulhon, and Marck}}]{Bonazzola:1998yq}
\bibinfo{author}{\bibfnamefont{S.}~\bibnamefont{Bonazzola}},
  \bibinfo{author}{\bibfnamefont{E.}~\bibnamefont{Gourgoulhon}},
  \bibnamefont{and} \bibinfo{author}{\bibfnamefont{J.-A.} \bibnamefont{Marck}},
  \bibinfo{journal}{Phys. Rev. Lett.} \textbf{\bibinfo{volume}{82}},
  \bibinfo{pages}{892} (\bibinfo{year}{1999}), \eprint{gr-qc/9810072}.

\bibitem[{\citenamefont{Marronetti et~al.}(1999)\citenamefont{Marronetti,
  Mathews, and Wilson}}]{Marronetti:1999ya}
\bibinfo{author}{\bibfnamefont{P.}~\bibnamefont{Marronetti}},
  \bibinfo{author}{\bibfnamefont{G.~J.} \bibnamefont{Mathews}},
  \bibnamefont{and} \bibinfo{author}{\bibfnamefont{J.~R.}
  \bibnamefont{Wilson}}, \bibinfo{journal}{Phys. Rev. D}
  \textbf{\bibinfo{volume}{60}}, \bibinfo{pages}{087301}
  (\bibinfo{year}{1999}), \eprint{arXiv:gr-qc/9906088}.

\bibitem[{\citenamefont{Uryu and Eriguchi}(2000)}]{Uryu:1999uu}
\bibinfo{author}{\bibfnamefont{K.}~\bibnamefont{Uryu}} \bibnamefont{and}
  \bibinfo{author}{\bibfnamefont{Y.}~\bibnamefont{Eriguchi}},
  \bibinfo{journal}{Phys. Rev.} \textbf{\bibinfo{volume}{D61}},
  \bibinfo{pages}{124023} (\bibinfo{year}{2000}), \eprint{gr-qc/9908059}.

\bibitem[{\citenamefont{Uryu et~al.}(2000)\citenamefont{Uryu, Shibata, and
  Eriguchi}}]{Uryu00a}
\bibinfo{author}{\bibfnamefont{K.}~\bibnamefont{Uryu}},
  \bibinfo{author}{\bibfnamefont{M.}~\bibnamefont{Shibata}}, \bibnamefont{and}
  \bibinfo{author}{\bibfnamefont{Y.}~\bibnamefont{Eriguchi}},
  \bibinfo{journal}{Phys. Rev.} \textbf{\bibinfo{volume}{D62}},
  \bibinfo{pages}{104015} (\bibinfo{year}{2000}).

\bibitem[{\citenamefont{Gourgoulhon et~al.}(2001)\citenamefont{Gourgoulhon,
  Grandclement, Taniguchi, Marck, and Bonazzola}}]{Gourgoulhon:2000nn}
\bibinfo{author}{\bibfnamefont{E.}~\bibnamefont{Gourgoulhon}},
  \bibinfo{author}{\bibfnamefont{P.}~\bibnamefont{Grandclement}},
  \bibinfo{author}{\bibfnamefont{K.}~\bibnamefont{Taniguchi}},
  \bibinfo{author}{\bibfnamefont{J.-A.} \bibnamefont{Marck}}, \bibnamefont{and}
  \bibinfo{author}{\bibfnamefont{S.}~\bibnamefont{Bonazzola}},
  \bibinfo{journal}{Phys. Rev.} \textbf{\bibinfo{volume}{D63}},
  \bibinfo{pages}{064029} (\bibinfo{year}{2001}), \eprint{gr-qc/0007028}.

\bibitem[{\citenamefont{Taniguchi et~al.}(2001)\citenamefont{Taniguchi,
  Gourgoulhon, and Bonazzola}}]{Taniguchi:2001qv}
\bibinfo{author}{\bibfnamefont{K.}~\bibnamefont{Taniguchi}},
  \bibinfo{author}{\bibfnamefont{E.}~\bibnamefont{Gourgoulhon}},
  \bibnamefont{and}
  \bibinfo{author}{\bibfnamefont{S.}~\bibnamefont{Bonazzola}},
  \bibinfo{journal}{Phys. Rev.} \textbf{\bibinfo{volume}{D64}},
  \bibinfo{pages}{064012} (\bibinfo{year}{2001}), \eprint{gr-qc/0103041}.

\bibitem[{\citenamefont{Taniguchi and Gourgoulhon}(2002)}]{Taniguchi:2002ns}
\bibinfo{author}{\bibfnamefont{K.}~\bibnamefont{Taniguchi}} \bibnamefont{and}
  \bibinfo{author}{\bibfnamefont{E.}~\bibnamefont{Gourgoulhon}},
  \bibinfo{journal}{Phys. Rev.} \textbf{\bibinfo{volume}{D66}},
  \bibinfo{pages}{104019} (\bibinfo{year}{2002}), \eprint{gr-qc/0207098}.

\bibitem[{\citenamefont{Kiuchi et~al.}(2009)\citenamefont{Kiuchi, Sekiguchi,
  Shibata, and Taniguchi}}]{Kiuchi:2009jt}
\bibinfo{author}{\bibfnamefont{K.}~\bibnamefont{Kiuchi}},
  \bibinfo{author}{\bibfnamefont{Y.}~\bibnamefont{Sekiguchi}},
  \bibinfo{author}{\bibfnamefont{M.}~\bibnamefont{Shibata}}, \bibnamefont{and}
  \bibinfo{author}{\bibfnamefont{K.}~\bibnamefont{Taniguchi}},
  \bibinfo{journal}{Phys.Rev.} \textbf{\bibinfo{volume}{D80}},
  \bibinfo{pages}{064037} (\bibinfo{year}{2009}), \eprint{0904.4551}.

\bibitem[{\citenamefont{Faber and Rasio}(2012)}]{Faber:2012rw}
\bibinfo{author}{\bibfnamefont{J.~A.} \bibnamefont{Faber}} \bibnamefont{and}
  \bibinfo{author}{\bibfnamefont{F.~A.} \bibnamefont{Rasio}},
  \bibinfo{journal}{Living Rev.Rel.} \textbf{\bibinfo{volume}{15}},
  \bibinfo{pages}{8} (\bibinfo{year}{2012}), \eprint{1204.3858}.

\bibitem[{\citenamefont{Uryu et~al.}(2006)\citenamefont{Uryu, Limousin,
  Friedman, Gourgoulhon, and Shibata}}]{Uryu:2005vv}
\bibinfo{author}{\bibfnamefont{K.}~\bibnamefont{Uryu}},
  \bibinfo{author}{\bibfnamefont{F.}~\bibnamefont{Limousin}},
  \bibinfo{author}{\bibfnamefont{J.~L.} \bibnamefont{Friedman}},
  \bibinfo{author}{\bibfnamefont{E.}~\bibnamefont{Gourgoulhon}},
  \bibnamefont{and} \bibinfo{author}{\bibfnamefont{M.}~\bibnamefont{Shibata}},
  \bibinfo{journal}{Phys.Rev.Lett.} \textbf{\bibinfo{volume}{97}},
  \bibinfo{pages}{171101} (\bibinfo{year}{2006}), \eprint{gr-qc/0511136}.

\bibitem[{\citenamefont{Cook and Baumgarte}(2008)}]{PhysRevD.78.104016}
\bibinfo{author}{\bibfnamefont{G.~B.} \bibnamefont{Cook}} \bibnamefont{and}
  \bibinfo{author}{\bibfnamefont{T.~W.} \bibnamefont{Baumgarte}},
  \bibinfo{journal}{Phys. Rev. D} \textbf{\bibinfo{volume}{78}},
  \bibinfo{pages}{104016} (\bibinfo{year}{2008}).

\bibitem[{\citenamefont{Uryu et~al.}(2009)\citenamefont{Uryu, Limousin,
  Friedman, Gourgoulhon, and Shibata}}]{Uryu:2009ye}
\bibinfo{author}{\bibfnamefont{K.}~\bibnamefont{Uryu}},
  \bibinfo{author}{\bibfnamefont{F.}~\bibnamefont{Limousin}},
  \bibinfo{author}{\bibfnamefont{J.~L.} \bibnamefont{Friedman}},
  \bibinfo{author}{\bibfnamefont{E.}~\bibnamefont{Gourgoulhon}},
  \bibnamefont{and} \bibinfo{author}{\bibfnamefont{M.}~\bibnamefont{Shibata}},
  \bibinfo{journal}{Phys.Rev.} \textbf{\bibinfo{volume}{D80}},
  \bibinfo{pages}{124004} (\bibinfo{year}{2009}), \eprint{0908.0579}.

\bibitem[{\citenamefont{Anderson et~al.}(2008)\citenamefont{Anderson,
  Hirschmann, Lehner, Liebling, Motl et~al.}}]{Anderson:2007kz}
\bibinfo{author}{\bibfnamefont{M.}~\bibnamefont{Anderson}},
  \bibinfo{author}{\bibfnamefont{E.~W.} \bibnamefont{Hirschmann}},
  \bibinfo{author}{\bibfnamefont{L.}~\bibnamefont{Lehner}},
  \bibinfo{author}{\bibfnamefont{S.~L.} \bibnamefont{Liebling}},
  \bibinfo{author}{\bibfnamefont{P.~M.} \bibnamefont{Motl}},
  \bibnamefont{et~al.}, \bibinfo{journal}{Phys.Rev.}
  \textbf{\bibinfo{volume}{D77}}, \bibinfo{pages}{024006}
  (\bibinfo{year}{2008}), \eprint{0708.2720}.

\bibitem[{\citenamefont{East et~al.}(2012)\citenamefont{East, Ramazanoglu, and
  Pretorius}}]{East:2012zn}
\bibinfo{author}{\bibfnamefont{W.~E.} \bibnamefont{East}},
  \bibinfo{author}{\bibfnamefont{F.~M.} \bibnamefont{Ramazanoglu}},
  \bibnamefont{and}
  \bibinfo{author}{\bibfnamefont{F.}~\bibnamefont{Pretorius}},
  \bibinfo{journal}{Phys.Rev.} \textbf{\bibinfo{volume}{D86}},
  \bibinfo{pages}{104053} (\bibinfo{year}{2012}), \eprint{1208.3473}.

\bibitem[{\citenamefont{Kastaun et~al.}(2013)\citenamefont{Kastaun, Galeazzi,
  Alic, Rezzolla, and Font}}]{Kastaun:2013mv}
\bibinfo{author}{\bibfnamefont{W.}~\bibnamefont{Kastaun}},
  \bibinfo{author}{\bibfnamefont{F.}~\bibnamefont{Galeazzi}},
  \bibinfo{author}{\bibfnamefont{D.}~\bibnamefont{Alic}},
  \bibinfo{author}{\bibfnamefont{L.}~\bibnamefont{Rezzolla}}, \bibnamefont{and}
  \bibinfo{author}{\bibfnamefont{J.~A.} \bibnamefont{Font}}
  (\bibinfo{year}{2013}), \eprint{1301.7348}.

\bibitem[{\citenamefont{Lyne et~al.}(2004)\citenamefont{Lyne, Burgay, Kramer,
  Possenti, Manchester et~al.}}]{Lyne:2004cj}
\bibinfo{author}{\bibfnamefont{A.}~\bibnamefont{Lyne}},
  \bibinfo{author}{\bibfnamefont{M.}~\bibnamefont{Burgay}},
  \bibinfo{author}{\bibfnamefont{M.}~\bibnamefont{Kramer}},
  \bibinfo{author}{\bibfnamefont{A.}~\bibnamefont{Possenti}},
  \bibinfo{author}{\bibfnamefont{R.}~\bibnamefont{Manchester}},
  \bibnamefont{et~al.}, \bibinfo{journal}{Science}
  \textbf{\bibinfo{volume}{303}}, \bibinfo{pages}{1153} (\bibinfo{year}{2004}),
  \eprint{astro-ph/0401086}.

\bibitem[{\citenamefont{Tichy}(2011)}]{Tichy:2011gw}
\bibinfo{author}{\bibfnamefont{W.}~\bibnamefont{Tichy}},
  \bibinfo{journal}{Phys.Rev.} \textbf{\bibinfo{volume}{D84}},
  \bibinfo{pages}{024041} (\bibinfo{year}{2011}).

\bibitem[{\citenamefont{Marronetti and Shapiro}(2003)}]{Marronetti:2003gk}
\bibinfo{author}{\bibfnamefont{P.}~\bibnamefont{Marronetti}} \bibnamefont{and}
  \bibinfo{author}{\bibfnamefont{S.~L.} \bibnamefont{Shapiro}},
  \bibinfo{journal}{Phys. Rev.} \textbf{\bibinfo{volume}{D68}},
  \bibinfo{pages}{104024} (\bibinfo{year}{2003}), \eprint{gr-qc/0306075}.

\bibitem[{\citenamefont{Baumgarte and Shapiro}(2009)}]{Baumgarte:2009fw}
\bibinfo{author}{\bibfnamefont{T.~W.} \bibnamefont{Baumgarte}}
  \bibnamefont{and} \bibinfo{author}{\bibfnamefont{S.~L.}
  \bibnamefont{Shapiro}}, \bibinfo{journal}{Phys.Rev.}
  \textbf{\bibinfo{volume}{D80}}, \bibinfo{pages}{064009}
  (\bibinfo{year}{2009}), \eprint{0909.0952}.

\bibitem[{\citenamefont{Tichy}(2012)}]{Tichy:2012rp}
\bibinfo{author}{\bibfnamefont{W.}~\bibnamefont{Tichy}},
  \bibinfo{journal}{Phys.Rev.} \textbf{\bibinfo{volume}{D86}},
  \bibinfo{pages}{064024} (\bibinfo{year}{2012}).

\bibitem[{LOR()}]{LORENE_web}
\urlprefix\url{http://www.lorene.obspm.fr/}.

\bibitem[{ET_()}]{ET_web}
\urlprefix\url{http://einsteintoolkit.org/}.

\bibitem[{\citenamefont{Loffler et~al.}(2012)\citenamefont{Loffler, Faber,
  Bentivegna, Bode, Diener, Haas, Hinder, Mundim, Ott, Schnetter
  et~al.}}]{ET2011}
\bibinfo{author}{\bibfnamefont{F.}~\bibnamefont{Loffler}},
  \bibinfo{author}{\bibfnamefont{J.}~\bibnamefont{Faber}},
  \bibinfo{author}{\bibfnamefont{E.}~\bibnamefont{Bentivegna}},
  \bibinfo{author}{\bibfnamefont{T.}~\bibnamefont{Bode}},
  \bibinfo{author}{\bibfnamefont{P.}~\bibnamefont{Diener}},
  \bibinfo{author}{\bibfnamefont{R.}~\bibnamefont{Haas}},
  \bibinfo{author}{\bibfnamefont{I.}~\bibnamefont{Hinder}},
  \bibinfo{author}{\bibfnamefont{B.~C.} \bibnamefont{Mundim}},
  \bibinfo{author}{\bibfnamefont{C.~D.} \bibnamefont{Ott}},
  \bibinfo{author}{\bibfnamefont{E.}~\bibnamefont{Schnetter}},
  \bibnamefont{et~al.}, \bibinfo{journal}{Class.Quant.Grav.}
  \textbf{\bibinfo{volume}{29}}, \bibinfo{pages}{115001}
  (\bibinfo{year}{2012}).

\bibitem[{\citenamefont{Campanelli et~al.}(2006)\citenamefont{Campanelli,
  Lousto, and Zlochower}}]{Campanelli:2006uy}
\bibinfo{author}{\bibfnamefont{M.}~\bibnamefont{Campanelli}},
  \bibinfo{author}{\bibfnamefont{C.O.}~\bibnamefont{Lousto}}, \bibnamefont{and}
  \bibinfo{author}{\bibfnamefont{Y.}~\bibnamefont{Zlochower}},
  \bibinfo{journal}{Phys.Rev.} \textbf{\bibinfo{volume}{D74}},
  \bibinfo{pages}{041501} (\bibinfo{year}{2006}), \eprint{gr-qc/0604012}.

\bibitem[{\citenamefont{Stergioulas}(2003)}]{lrr-2003-3}
\bibinfo{author}{\bibfnamefont{N.}~\bibnamefont{Stergioulas}},
  \bibinfo{journal}{Living Reviews in Relativity} \textbf{\bibinfo{volume}{6}}
  (\bibinfo{year}{2003}).

\bibitem[{\citenamefont{Stergioulas and Friedman}(1995)}]{Stergioulas95}
\bibinfo{author}{\bibfnamefont{N.}~\bibnamefont{Stergioulas}} \bibnamefont{and}
  \bibinfo{author}{\bibfnamefont{J.~L.} \bibnamefont{Friedman}},
  \bibinfo{journal}{Astrophys. J.} \textbf{\bibinfo{volume}{444}},
  \bibinfo{pages}{306} (\bibinfo{year}{1995}).

\bibitem[{RNS()}]{RNS_web}
\urlprefix\url{http://www.gravity.phys.uwm.edu/rns/}.

\bibitem[{\citenamefont{Komatsu et~al.}(1989)\citenamefont{Komatsu, Eriguchi,
  and Hachisu}}]{Komatsu89}
\bibinfo{author}{\bibfnamefont{H.}~\bibnamefont{Komatsu}},
  \bibinfo{author}{\bibfnamefont{Y.}~\bibnamefont{Eriguchi}}, \bibnamefont{and}
  \bibinfo{author}{\bibfnamefont{I.}~\bibnamefont{Hachisu}},
  \bibinfo{journal}{Mon. Not. R. Astron. Soc.} \textbf{\bibinfo{volume}{237}},
  \bibinfo{pages}{355} (\bibinfo{year}{1989}).

\bibitem[{\citenamefont{{Cook} et~al.}(1994)\citenamefont{{Cook}, {Shapiro},
  and {Teukolsky}}}]{Cook1994ApJ}
\bibinfo{author}{\bibfnamefont{G.~B.} \bibnamefont{{Cook}}},
  \bibinfo{author}{\bibfnamefont{S.~L.} \bibnamefont{{Shapiro}}},
  \bibnamefont{and} \bibinfo{author}{\bibfnamefont{S.~A.}
  \bibnamefont{{Teukolsky}}}, \bibinfo{journal}{\apj}
  \textbf{\bibinfo{volume}{422}}, \bibinfo{pages}{227} (\bibinfo{year}{1994}).

\bibitem[{\citenamefont{Matzner et~al.}(1998)\citenamefont{Matzner, Huq, and
  Shoemaker}}]{Matzner98a}
\bibinfo{author}{\bibfnamefont{R.~A.} \bibnamefont{Matzner}},
  \bibinfo{author}{\bibfnamefont{M.~F.} \bibnamefont{Huq}}, \bibnamefont{and}
  \bibinfo{author}{\bibfnamefont{D.}~\bibnamefont{Shoemaker}},
  \bibinfo{journal}{Phys. Rev. D} \textbf{\bibinfo{volume}{59}},
  \bibinfo{pages}{024015} (\bibinfo{year}{1998}).

\bibitem[{\citenamefont{Marronetti et~al.}(2000)\citenamefont{Marronetti, Huq,
  Laguna, Lehner, Matzner, and Shoemaker}}]{Marronetti00a}
\bibinfo{author}{\bibfnamefont{P.}~\bibnamefont{Marronetti}},
  \bibinfo{author}{\bibfnamefont{M.~F.} \bibnamefont{Huq}},
  \bibinfo{author}{\bibfnamefont{P.}~\bibnamefont{Laguna}},
  \bibinfo{author}{\bibfnamefont{L.}~\bibnamefont{Lehner}},
  \bibinfo{author}{\bibfnamefont{R.~A.} \bibnamefont{Matzner}},
  \bibnamefont{and}
  \bibinfo{author}{\bibfnamefont{D.}~\bibnamefont{Shoemaker}},
  \bibinfo{journal}{Phys. Rev. D} \textbf{\bibinfo{volume}{62}},
  \bibinfo{pages}{024017} (\bibinfo{year}{2000}),
  \bibinfo{note}{gr-qc/0001077}.

\bibitem[{\citenamefont{Schnetter et~al.}(2005)\citenamefont{Schnetter,
  Herrmann, and Pollney}}]{Schnetter04}
\bibinfo{author}{\bibfnamefont{E.}~\bibnamefont{Schnetter}},
  \bibinfo{author}{\bibfnamefont{F.}~\bibnamefont{Herrmann}}, \bibnamefont{and}
  \bibinfo{author}{\bibfnamefont{D.}~\bibnamefont{Pollney}},
  \bibinfo{journal}{Phys. Rev. D} \textbf{\bibinfo{volume}{71}},
  \bibinfo{pages}{044033} (\bibinfo{year}{2005}), \eprint{gr-qc/0410081}.

\bibitem[{\citenamefont{Nakamura et~al.}(1987)\citenamefont{Nakamura, Oohara,
  and Kojima}}]{Nakamura87}
\bibinfo{author}{\bibfnamefont{T.}~\bibnamefont{Nakamura}},
  \bibinfo{author}{\bibfnamefont{K.}~\bibnamefont{Oohara}}, \bibnamefont{and}
  \bibinfo{author}{\bibfnamefont{Y.}~\bibnamefont{Kojima}},
  \bibinfo{journal}{Prog. Theor. Phys. Suppl.} \textbf{\bibinfo{volume}{90}},
  \bibinfo{pages}{1} (\bibinfo{year}{1987}).

\bibitem[{\citenamefont{Shibata and Nakamura}(1995)}]{Shibata95}
\bibinfo{author}{\bibfnamefont{M.}~\bibnamefont{Shibata}} \bibnamefont{and}
  \bibinfo{author}{\bibfnamefont{T.}~\bibnamefont{Nakamura}},
  \bibinfo{journal}{Phys. Rev. D} \textbf{\bibinfo{volume}{52}},
  \bibinfo{pages}{5428} (\bibinfo{year}{1995}).

\bibitem[{\citenamefont{Baumgarte and Shapiro}(1998)}]{Baumgarte:1998te}
\bibinfo{author}{\bibfnamefont{T.~W.} \bibnamefont{Baumgarte}}
  \bibnamefont{and} \bibinfo{author}{\bibfnamefont{S.~L.}
  \bibnamefont{Shapiro}}, \bibinfo{journal}{Phys. Rev.}
  \textbf{\bibinfo{volume}{D59}}, \bibinfo{pages}{024007}
  (\bibinfo{year}{1998}), \eprint{gr-qc/9810065}.

\bibitem[{\citenamefont{Brown et~al.}(2009)\citenamefont{Brown, Diener,
  Sarbach, Schnetter, and Tiglio}}]{Brown:2008sb}
\bibinfo{author}{\bibfnamefont{D.} \bibnamefont{Brown}},
  \bibinfo{author}{\bibfnamefont{P.}~\bibnamefont{Diener}},
  \bibinfo{author}{\bibfnamefont{O.}~\bibnamefont{Sarbach}},
  \bibinfo{author}{\bibfnamefont{E.}~\bibnamefont{Schnetter}},
  \bibnamefont{and} \bibinfo{author}{\bibfnamefont{M.}~\bibnamefont{Tiglio}},
  \bibinfo{journal}{Phys.Rev.} \textbf{\bibinfo{volume}{D79}},
  \bibinfo{pages}{044023} (\bibinfo{year}{2009}), \eprint{0809.3533}.

\bibitem[{\citenamefont{Baiotti et~al.}(2005)\citenamefont{Baiotti, Hawke,
  Montero, L{\"o}ffler, Rezzolla, Stergioulas, Font, and Seidel}}]{Baiotti04a}
\bibinfo{author}{\bibfnamefont{L.}~\bibnamefont{Baiotti}},
  \bibinfo{author}{\bibfnamefont{I.}~\bibnamefont{Hawke}},
  \bibinfo{author}{\bibfnamefont{P.~J.} \bibnamefont{Montero}},
  \bibinfo{author}{\bibfnamefont{F.}~\bibnamefont{L{\"o}ffler}},
  \bibinfo{author}{\bibfnamefont{L.}~\bibnamefont{Rezzolla}},
  \bibinfo{author}{\bibfnamefont{N.}~\bibnamefont{Stergioulas}},
  \bibinfo{author}{\bibfnamefont{J.~A.} \bibnamefont{Font}}, \bibnamefont{and}
  \bibinfo{author}{\bibfnamefont{E.}~\bibnamefont{Seidel}},
  \bibinfo{journal}{Phys. Rev. D} \textbf{\bibinfo{volume}{71}},
  \bibinfo{pages}{024035} (\bibinfo{year}{2005}), \eprint{gr-qc/0403029}.

\bibitem[{\citenamefont{Tichy and Marronetti}(2011)}]{Tichy:2010qa}
\bibinfo{author}{\bibfnamefont{W.}~\bibnamefont{Tichy}} \bibnamefont{and}
  \bibinfo{author}{\bibfnamefont{P.}~\bibnamefont{Marronetti}},
  \bibinfo{journal}{Phys.Rev.} \textbf{\bibinfo{volume}{D83}},
  \bibinfo{pages}{024012} (\bibinfo{year}{2011}), \eprint{1010.2936}.

\bibitem[{\citenamefont{Husa et~al.}(2008)\citenamefont{Husa, Hannam, Gonzalez,
  Sperhake, and Brugmann}}]{Husa:2007rh}
\bibinfo{author}{\bibfnamefont{S.}~\bibnamefont{Husa}},
  \bibinfo{author}{\bibfnamefont{M.}~\bibnamefont{Hannam}},
  \bibinfo{author}{\bibfnamefont{J.~A.} \bibnamefont{Gonzalez}},
  \bibinfo{author}{\bibfnamefont{U.}~\bibnamefont{Sperhake}}, \bibnamefont{and}
  \bibinfo{author}{\bibfnamefont{B.}~\bibnamefont{Brugmann}},
  \bibinfo{journal}{Phys.Rev.} \textbf{\bibinfo{volume}{D77}},
  \bibinfo{pages}{044037} (\bibinfo{year}{2008}), \eprint{0706.0904}.

\bibitem[{\citenamefont{Walther et~al.}(2009)\citenamefont{Walther, Brugmann,
  and Mueller}}]{Walther:2009ng}
\bibinfo{author}{\bibfnamefont{B.}~\bibnamefont{Walther}},
  \bibinfo{author}{\bibfnamefont{B.}~\bibnamefont{Brugmann}},
  \bibnamefont{and} \bibinfo{author}{\bibfnamefont{D.}~\bibnamefont{Muller}},
  \bibinfo{journal}{Phys.Rev.} \textbf{\bibinfo{volume}{D79}},
  \bibinfo{pages}{124040} (\bibinfo{year}{2009}), \eprint{0901.0993}.

\bibitem[{\citenamefont{Pfeiffer et~al.}(2007)\citenamefont{Pfeiffer, Brown,
  Kidder, Lindblom, Lovelace et~al.}}]{Pfeiffer:2007yz}
\bibinfo{author}{\bibfnamefont{H.~P.} \bibnamefont{Pfeiffer}},
  \bibinfo{author}{\bibfnamefont{D.~A.} \bibnamefont{Brown}},
  \bibinfo{author}{\bibfnamefont{L.~E.} \bibnamefont{Kidder}},
  \bibinfo{author}{\bibfnamefont{L.}~\bibnamefont{Lindblom}},
  \bibinfo{author}{\bibfnamefont{G.}~\bibnamefont{Lovelace}},
  \bibnamefont{et~al.}, \bibinfo{journal}{Class.Quant.Grav.}
  \textbf{\bibinfo{volume}{24}}, \bibinfo{pages}{S59} (\bibinfo{year}{2007}),
  \eprint{gr-qc/0702106}.

\bibitem[{\citenamefont{Boyle et~al.}(2007)\citenamefont{Boyle, Brown, Kidder,
  Mroue, Pfeiffer et~al.}}]{Boyle:2007ft}
\bibinfo{author}{\bibfnamefont{M.}~\bibnamefont{Boyle}},
  \bibinfo{author}{\bibfnamefont{D.~A.} \bibnamefont{Brown}},
  \bibinfo{author}{\bibfnamefont{L.~E.} \bibnamefont{Kidder}},
  \bibinfo{author}{\bibfnamefont{A.~H.} \bibnamefont{Mroue}},
  \bibinfo{author}{\bibfnamefont{H.~P.} \bibnamefont{Pfeiffer}},
  \bibnamefont{et~al.}, \bibinfo{journal}{Phys.Rev.}
  \textbf{\bibinfo{volume}{D76}}, \bibinfo{pages}{124038}
  (\bibinfo{year}{2007}), \eprint{0710.0158}.

\bibitem[{\citenamefont{Mroue et~al.}(2010)\citenamefont{Mroue, Pfeiffer,
  Kidder, and Teukolsky}}]{Mroue:2010re}
\bibinfo{author}{\bibfnamefont{A.~H.} \bibnamefont{Mroue}},
  \bibinfo{author}{\bibfnamefont{H.~P.} \bibnamefont{Pfeiffer}},
  \bibinfo{author}{\bibfnamefont{L.~E.} \bibnamefont{Kidder}},
  \bibnamefont{and} \bibinfo{author}{\bibfnamefont{S.~A.}
  \bibnamefont{Teukolsky}}, \bibinfo{journal}{Phys.Rev.}
  \textbf{\bibinfo{volume}{D82}}, \bibinfo{pages}{124016}
  (\bibinfo{year}{2010}), \eprint{1004.4697}.

\bibitem[{\citenamefont{Kidder}(1995)}]{Kidder:1995zr}
\bibinfo{author}{\bibfnamefont{L.~E.} \bibnamefont{Kidder}},
  \bibinfo{journal}{Phys. Rev.} \textbf{\bibinfo{volume}{D52}},
  \bibinfo{pages}{821} (\bibinfo{year}{1995}), \eprint{gr-qc/9506022}.

\bibitem[{\citenamefont{Bernuzzi and Hilditch}(2010)}]{Bernuzzi:2009ex}
\bibinfo{author}{\bibfnamefont{S.}~\bibnamefont{Bernuzzi}} \bibnamefont{and}
  \bibinfo{author}{\bibfnamefont{D.}~\bibnamefont{Hilditch}},
  \bibinfo{journal}{Phys.Rev.} \textbf{\bibinfo{volume}{D81}},
  \bibinfo{pages}{084003} (\bibinfo{year}{2010}), \eprint{0912.2920}.

\bibitem[{\citenamefont{Alic et~al.}(2012)\citenamefont{Alic, Bona-Casas, Bona,
  Rezzolla, and Palenzuela}}]{Alic:2011gg}
\bibinfo{author}{\bibfnamefont{D.}~\bibnamefont{Alic}},
  \bibinfo{author}{\bibfnamefont{C.}~\bibnamefont{Bona-Casas}},
  \bibinfo{author}{\bibfnamefont{C.}~\bibnamefont{Bona}},
  \bibinfo{author}{\bibfnamefont{L.}~\bibnamefont{Rezzolla}}, \bibnamefont{and}
  \bibinfo{author}{\bibfnamefont{C.}~\bibnamefont{Palenzuela}},
  \bibinfo{journal}{Phys.Rev.} \textbf{\bibinfo{volume}{D85}},
  \bibinfo{pages}{064040} (\bibinfo{year}{2012}), \eprint{1106.2254}.

\bibitem[{\citenamefont{Hilditch et~al.}(2012)\citenamefont{Hilditch, Bernuzzi,
  Thierfelder, Cao, Tichy et~al.}}]{Hilditch:2012fp}
\bibinfo{author}{\bibfnamefont{D.}~\bibnamefont{Hilditch}},
  \bibinfo{author}{\bibfnamefont{S.}~\bibnamefont{Bernuzzi}},
  \bibinfo{author}{\bibfnamefont{M.}~\bibnamefont{Thierfelder}},
  \bibinfo{author}{\bibfnamefont{Z.}~\bibnamefont{Cao}},
  \bibinfo{author}{\bibfnamefont{W.}~\bibnamefont{Tichy}}, \bibnamefont{et~al.}
  (\bibinfo{year}{2012}), \eprint{1212.2901}.

\bibitem[{\citenamefont{Alic et~al.}(2013)\citenamefont{Alic, Kastaun, and
  Rezzolla}}]{Alic:2013xsa}
\bibinfo{author}{\bibfnamefont{D.}~\bibnamefont{Alic}},
  \bibinfo{author}{\bibfnamefont{W.}~\bibnamefont{Kastaun}}, \bibnamefont{and}
  \bibinfo{author}{\bibfnamefont{L.}~\bibnamefont{Rezzolla}}
  (\bibinfo{year}{2013}), \eprint{1307.7391}.

\end{thebibliography}

\end{document}